\documentclass[useAMS]{mn2e}
\usepackage{epsfig,natbib_vw,times}

\bibpunct[, ]{(}{)}{;}{a}{}{,}

\def \mnras {MNRAS}

\def \be {\begin{equation}}
\def \ee {\end{equation}}

\def\gsim{\mathrel{\lower0.6ex\hbox{$\buildrel {\textstyle >}
 \over {\scriptstyle \sim}$}}}
\def\lsim{\mathrel{\lower0.6ex\hbox{$\buildrel {\textstyle <}
 \over {\scriptstyle \sim}$}}}

\def\m@th{\mathsurround=0pt }
\def\eqalign#1{\null\,\vcenter{\openup1\jot \m@th
 \ialign{\strut\hfil$\displaystyle{##}$&$\displaystyle{{}##}$\hfil
 \crcr#1\crcr}}\,}

\def \caii {Ca\,{\sc ii}~}

\def \oii {[O\,{\sc ii}]~}
\def \oiii {[O\,{\sc iii}]~}
\def \nv {N\,{\sc v}~}

\def \mgii {Mg\,{\sc ii}~}

\def \civ {C\,{\sc iv}~}
\def \ciii {C\,{\sc iii}]~}
\def \aliii {Al\,{\sc iii}~}
\def \siiii {Si\,{\sc iii}]~}

\def \ciiicc {C\,{\sc iii}]\_cc~}
\def \civcc {C\,{\sc iv}\_cc~}
\def \mgiicc {Mg\,{\sc ii}\_cc~}

\def \kms {\rm \,km\,s$^{-1}$}
\title[Improved SDSS redshifts]{Improved redshifts for SDSS quasar spectra}
\author[P. C. Hewett and V. Wild ]{Paul C. Hewett$^{1}$\thanks{phewett@ast.cam.ac.uk}
  and Vivienne Wild$^2$
\vspace*{6pt}\\
1. Institute of Astronomy, University of Cambridge, Cambridge CB3 0HA, UK \\
2. Institut d'Astrophysique de Paris, 98bis Boulevard Arago, 75014 Paris, France\\}

\begin{document}

\maketitle
\begin{abstract}

A systematic investigation of the relationship between different redshift
estimation schemes for more than 91\,000 quasars in the Sloan Digital Sky Survey
(SDSS) Data Release 6 (DR6) is presented.  The publicly available SDSS quasar
redshifts are shown to possess systematic biases of $\Delta z/(1+z)$$\ge$0.002
(600\kms) over both small ($\delta z$$\simeq$0.1) and large ($\delta
z$$\simeq$1) redshift intervals.  Empirical relationships between redshifts
based on i) \caii H \& K host galaxy absorption, ii) quasar \oii $\lambda$3728,
iii) \oiii $\lambda\lambda$4960,5008 emission, and iv) cross-correlation (with a
master quasar template) that includes, at increasing quasar redshift, the
prominent \mgii $\lambda$2799, \ciii $\lambda$1908 and \civ $\lambda$1549
emission lines, are established as a function of quasar redshift and luminosity.
New redshifts in the resulting catalogue possess systematic biases a factor of
$\simeq$20 lower compared to the SDSS redshift values; systematic effects are
reduced to the level of $\Delta z/(1+z)$$\le$10$^{-4}$ (30\kms) per unit
redshift, or $\le$2.5$\times$10$^{-5}$ per unit absolute magnitude.  Redshift
errors, including components due both to internal reproducibility and
the intrinsic quasar-to-quasar variation among the population, are available for
all quasars in the catalogue.  The improved redshifts and their associated
errors have wide applicability in areas such as quasar absorption outflows,
quasar clustering, quasar-galaxy clustering and proximity-effect determinations.

\end{abstract}


\section{Introduction}\label{sec:intro}

The Sloan Digital Sky Survey (SDSS) \citep{2000AJ....120.1579Y} has produced a
revolution in both the volume and quality of spectroscopic data available for
quasars.  The Data Release 5 (DR5) \citep{2007ApJS..172..634A} and Legacy Data
Release 7 (DR7) \citep{2009ApJS..182..543A} with their associated quasar
catalogues \citep[respectively]{2007AJ....134..102S, 2010AJ....XXX..XXXX}
provide intermediate resolution ($R$$\sim$2000), moderate signal-to-noise ratio
(SNR) (SNR$\sim$15 per 69\kms pixel), spectra of unprecedented homogeneity,
covering essentially the entire ''optical'' wavelength region
($\lambda$=3800--9180\,\AA).

The quality of the Schneider et al.  quasar catalogues is truly impressive, with
errors in redshift identification reduced to the 0.01 per cent level and
individual redshift estimates, resulting primarily from the SDSS spectroscopic
pipeline \citep[and the SDSS DR7
website\footnote{http://www.sdss.org/dr7/algorithms/redshift\_type.html}]{2002AJ....123..485S},
are accurate to of order $\Delta z/(1+z)$$\sim$0.002.  The publication of even
individual quasar redshifts, based on moderate resolution spectra, to such
accuracy was a significant achievement prior to the mid-1990s, further
highlighting the advance represented by the SDSS.

Notwithstanding the quality of the SDSS quasar spectra and the associated
redshift estimates, important scientific investigations, including the
clustering of quasars themselves \citep[e.g.][]{2002MNRAS.335..459C,
2007AJ....133.2222S}, the cross-correlation of quasars and other object
populations \citep[e.g.][]{2009MNRAS.397.1862P}, the proximity effect
\citep[e.g.][]{1988ApJ...327..570B, 2008MNRAS.391.1457K}, the origin and
properties of associated absorbers \citep[e.g.][]{2008MNRAS.386.2055N,
2008MNRAS.388..227W, 2009MNRAS.392.1539T} benefit significantly both from
reduced systematics in redshift determinations and the reliable assignment of
redshift uncertainties for individual quasars.

In this paper we present the determination of new redshifts and associated error
estimates for more than 89\,500 quasars from the SDSS DR6
\citep{2008ApJS..175..297A}.  Our redshift determinations suffer from much
smaller systematic effects compared to the default values from the SDSS
spectroscopic pipeline.  Specifically, systematics are reduced by more than an
order of magnitude to 1.0$\times$10$^{-4}$ in $\Delta z$/(1+$z$) per unit
redshift, or, equivalently, 30\kms per unit redshift\footnote{The quasar
research community has normally quantified redshift errors in terms of $\Delta
z$/(1+$z$), whereas researchers studying galaxies conventionally specify
uncertainties in kilometres per second.  The improvements possible in redshift
determination made possible by the SDSS spectra are such that the `kilometres
per second' parameterisation is increasingly attractive and we specify the main
results using both schemes.}.  A detailed comparison of redshifts derived from
\caii H\&K absorption, \oii $\lambda\lambda$3727,3729 emission, \oiii
$\lambda\lambda$4960,5008 emission and cross-correlation with a new quasar
template spectrum, provides greatly improved error estimates for individual
quasar redshifts.  The error estimates incorporate both the uncertainties
resulting from the properties of the SDSS spectra, quantified using the very
large number of multiple spectra present in the SDSS, and the intrinsic
quasar-to-quasar dispersion.  The resulting catalogue will allow significant
advances in many studies that rely on the determination of systemic quasar
redshifts with small systematics and well-determined uncertainties.

The paper is structured as follows.  Section 2 describes the quasar sample,
before the features of the quasar redshifts available from the SDSS
spectroscopic pipeline are illustrated in Section 3.  Section 4 includes a
description of the procedures involved in generating a master quasar template
for the cross-correlation redshift estimates.  Section 5 then describes the
procedures employed to provide the new redshift estimates for the SDSS quasars.
An assessment of the consistency of the different redshift estimates is given in
Section 6 and redshift estimates, based on different rest-frame wavelength
regions, are placed onto the same `systemic' reference system.  A critical
assessment of the internal and external reliability of the new quasar redshifts
is presented at this point.  Twenty-one centimetre radio observations of the
majority of the SDSS quasars are available from the Faint Images of the Radio
Sky at Twenty centimetres \citep[FIRST,][]{1995ApJ...450..559B}.  Section 7
contains a description showing how the new redshift estimation scheme allows
spectral energy distribution (SED) dependent composite spectra (for
FIRST-detected quasars in this case) to be constructed, producing significantly
improved redshifts.  The resulting redshift catalogue, including well-determined
error estimates for each quasar, is described in Section 8.  A short discussion,
including consideration of the origin of the differences with published
redshifts and an independent test of the new redshifts follows in Section 9.
The paper concludes with a brief summary of the conclusions as Section 10.  We
adopt the same convention as employed in the SDSS and use vacuum wavelengths
throughout the paper. Absolute magnitudes are calculated in a cosmology with
$H_0$=70\kms, $\Omega_M$=0.3 and $\Omega_\Lambda$=0.7.

\section{Quasar Sample}\label{sec:samp}

The quasar sample consists of 91\,665 objects, including 77\,392 quasars in the
\citet{2007AJ....134..102S} DR5 catalogue that are retained in the later
DR7 quasar catalogue of \cite{2010AJ....XXX..XXXX}.  A further 13\,081 objects
are quasars, present in the additional DR6 spectroscopic plates, identified by
one of us (PCH) using a similar prescription to that employed by
\citet{2007AJ....134..102S}, all of which are present in the
\citet{2010AJ....XXX..XXXX} catalogue.  An additional 1192 objects, which do not satisfy
one, or both, of the emission line velocity width or absolute magnitude
criterion imposed by \citet{2007AJ....134..102S}, are also included.  While
formally failing the `quasar' definition of Schneider et al.'s DR5 and DR7
compilations the objects are essentially all luminous active galactic nuclei
(AGN).  None of the results in the paper depend on the exact definition of the
`quasar'-sample used.

The spectra were all processed through the sky-residual subtraction scheme of
\citet{2005MNRAS.358.1083W}, resulting in significantly improved SNR at
wavelengths $\lambda$$>$7200\,\AA.  The SNR improvement allows the important
quasar rest-frame wavelength regions containing the \mgii $\lambda$2799 and
\ciii $\lambda$1908 emission lines, to contribute to the cross-correlation
redshift determinations (Section \ref{sec:cc_calc}) to much higher redshifts
than is possible using the original SDSS spectra.

The SDSS DR6 contains a very large number of objects for which multiple spectra
are available.  For our quasar sample there are $\simeq$9000 independent pairs
of spectra.  The catalogue of spectrum pairs allows the accurate
determination of redshift reproducibility as a function of SNR, redshift and
cross-correlation amplitude and extensive use of the spectrum pairs is made to
quantify the contribution of the SDSS spectra themselves to the quasar redshift
errors.

\section{SDSS Redshifts}\label{sec:sdss_zs}

The SDSS spectroscopic pipeline ({\tt spectro1d}) incorporates a sophisticated
scheme$^1$ for determining both the classification (star, galaxy, quasar,...)
of the spectra and the redshifts of extragalactic objects.  Cross-correlation
redshift estimates are determined using the \citet{1979AJ.....84.1511T}
technique and a composite quasar template from \citep{2001AJ....122..549V}.
Emission lines are identified via a wavelet transform technique and an
independent redshift estimate is derived using the observed-frame wavelength
emission line locations and reference rest-frame emission line wavelengths, the
latter taken from the \citet{2001AJ....122..549V} composite quasar spectrum.
The reference
wavelengths\footnote{http://www.sdss.org/dr7/dm/flatFiles/spSpec.html describes
the {\tt spectro1d} FITS-file data model and lists the emission line
wavelengths.} adopted from the quasar composite can differ from laboratory values
due to the complex, often asymmetric, line profiles and apparent `velocity
shifts' of the line centroids \citep{1982ApJ...263...79G, 1992ApJS...79....1T,
2002AJ....124....1R}.

The SDSS database and the individual FITS spectrum files contain
extensive quantitative information on the determination and
reliability of the different redshift estimates.  However, the
majority of researchers utilise the `final'-redshift estimate {\tt z}
included in the SDSS {\tt SpecObjAll} table, the individual FITS
spectrum file headers, or from the Schneider et al.
catalogues\footnote{In the Schneider et al. DR5 and DR7 quasar
catalogues, catastrophic redshift errors are virtually absent but
otherwise the catalogued redshifts are the `final'-redshift estimates
from the pipeline reductions.}.

If available, the cross-correlation redshift is adopted as the
`final' redshift for the spectrum.  Some 88 per cent of the quasars
possess redshifts derived from cross-correlation and more than a
third of such objects also possess consistent emission line-based
redshifts.  A further 7 per cent of quasars, where no reliable
cross-correlation redshift is available, possess redshifts derived
from the emission lines. The remaining 5 per cent of quasars,
including a large fraction of pathological objects and spectra of low
SNR, have redshifts derived via manual inspection of the spectra.

\subsection{SDSS Princeton redshifts}

Independent spectrum classifications and redshift determinations,
based on direct $\chi^2$-fitting of template spectra to the data,
have been made at Princeton using the {\tt specBS}
code\footnote{http://spectro.princeton.edu/}.  The redshift
determination, essentially via cross-correlation, differs from the
implementation employed in the {\tt spectro1d} pipeline but the same
composite quasar template from \citet{2001AJ....122..549V} was used.

\subsection{SDSS Redshift intercomparison}

\begin{figure}
 \begin{minipage}{\textwidth}
\includegraphics[scale=0.30,angle=-90]{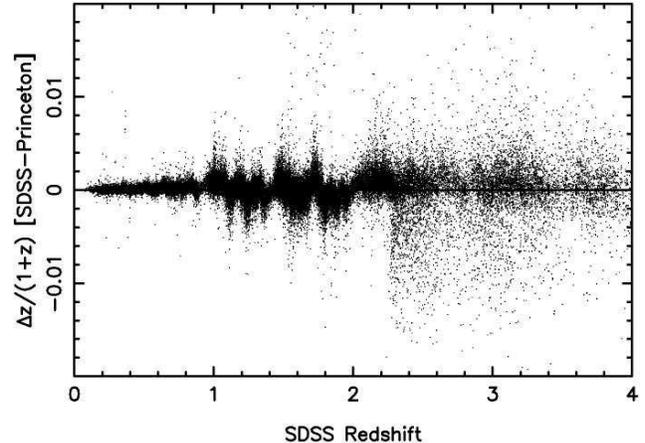}
  \end{minipage}
\caption{Redshift differences, $\Delta z/(1+z)$, between SDSS and Princeton
pipeline reductions.  Spectra plotted possess SDSS final-redshifts with
high-confidence (zConf$>$0.9) derived via cross-correlation (zstatus=3 or 4).
Large differences between redshifts extend to amplitudes of
$\pm$5$\times$10$^{-3}$ ($\pm$1500\kms).  Particularly striking is the sequence
of apparent discontinuities in the behaviour of the redshift differences as a
function of redshift. Data for 69\,915 spectra are included (503 lie outside the y-axis range
plotted).}
\label{sdss_prince}
\end{figure}

Fig.~\ref{sdss_prince} shows a comparison of the SDSS final-redshifts
and Princeton redshifts as a function of quasar redshift\footnote{All
figures showing redshift differences between estimates $z_1$ and $z_2$
have $\Delta z/(1+z)$=$(z_1-z_2)/(1+z)$ plotted as the y-axis. The
choice of which estimate is used in the denominator is usually
irrelevant given the scale of the plots.}.  The selection of the
sub-sample of more than 70\,000 spectra is conservative in that only
spectra with high-confidence SDSS redshifts, where there is also no
inconsistency between the cross-correlation and emission line
redshift determinations, are used. The data in Fig.~\ref{sdss_prince} 
should essentially represent an internal consistency check and the large
differences between redshifts, extending to $\pm$5$\times$10$^{-3}$,
or 1500\kms, are surprising. Perhaps even more striking is the sequence
of apparent discontinuities in the behaviour as a function of redshift.

\begin{figure}
 \begin{minipage}{\textwidth}
  \includegraphics[scale=0.30,angle=-90]{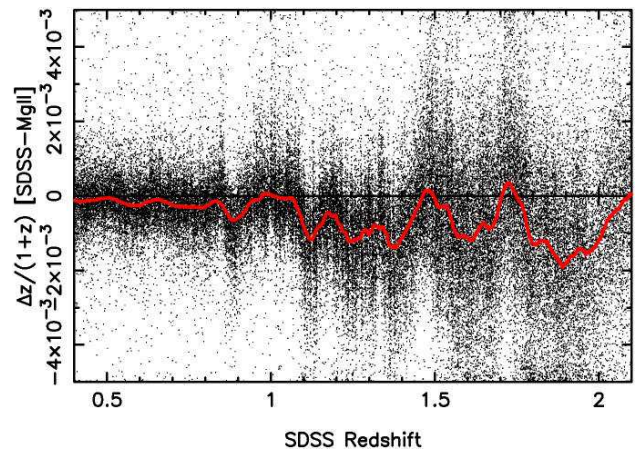}
  \end{minipage}
  \caption{Redshift differences,$\Delta z/(1+z)$, between SDSS redshifts and
redshifts derived from the SDSS-determined \mgii emission line locations.
Spectra plotted possess \mgii emission line SNR$>$10.  The solid red line,
calculated using a 2001-point running median of the data-points, shows
the form of the systematic trends with redshift. Systematic redshift 
differences of $\simeq$500\kms shifts over small redshift intervals are
evident and, over a larger redshift interval, a prominent 
systematic trend of 2$\times$10$^{-3}$ (600\kms) can be seen.
Data for 60\,190 spectra are included (2101 lie outside the y-axis range 
plotted). }
\label{sdss_mgii}
\end{figure}

\begin{figure}
 \begin{minipage}{\textwidth}
  \includegraphics[scale=0.30,angle=-90]{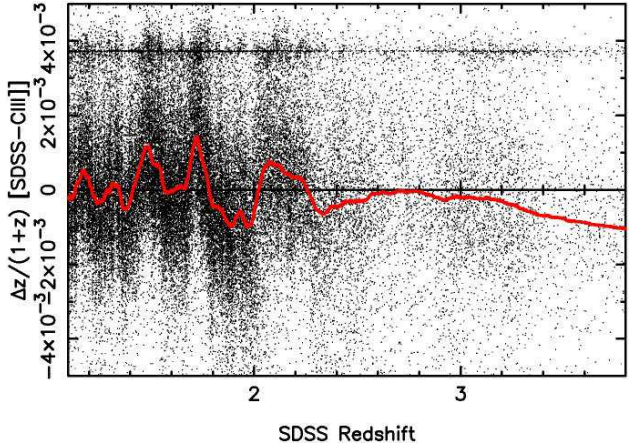}
  \end{minipage}
 \caption{Redshift differences,$\Delta z/(1+z)$, between SDSS redshifts
and redshifts derived from the SDSS-determined \ciii emission line
locations.  Spectra plotted possess \ciii emission line SNR$>$10.  The
wavelength adopted for the \ciii emission centroid has been chosen to
produce, on average, zero-offset for $z$$<$2.2.  The solid red line,
calculated using a 2001-point running median of the data-points, shows the
form of the systematic trends with redshift.  Large systematic changes of
up to 3$\times$10$^{-3}$ (900\kms) over small redshift intervals can be
seen.  The prominent horizontal `feature' close to
$\simeq$4$\times$10$^{-3}$ is an artifact resulting from the $\pm$1500\kms
velocity interval (about the cross-correlation redshift) within which the
SDSS reduction pipeline searches for emission lines.  Data for 59\,260
spectra are included (2628 lie outside the y-axis range plotted).}
\label{sdss_ciii}
\end{figure}

A second illustration of the extent of redshift-dependent systematics comes from
comparing the redshift derived from the location of the \mgii
$\lambda\lambda$2796,2803 emission in each quasar spectrum with the SDSS
redshift.  Fig.~\ref{sdss_mgii} presents the data for more than 60\,000 spectra
with SNR$\ge$10 \mgii emission line locations (from the SDSS spectroscopic
pipeline\footnote{The SNR constraints applied to the use of emission lines
refer to the significance of the emission line detection by the SDSS
spectroscopic pipeline.}).  The rest-frame location of the \mgii emission line
has been shown by many studies over the decades to be well-behaved and there is
no reason to expect $\simeq$500\kms shifts over small redshift intervals, or,
indeed, an apparent systematic 2$\times$10$^{-3}$ (600\kms) change in the
location of the \mgii emission with increasing redshift of the quasars.  The
systematic redshift differences show similar patterns over the redshift range
common to both Fig.~\ref{sdss_prince} and Fig.~\ref{sdss_mgii}.  Although
somewhat more complex to interpret (Section~\ref{sec:em_shifts}), the equivalent
plot for the \ciii emission, Fig.~\ref{sdss_ciii}, also shows strong systematic
effects as a function of redshift.  The form and substantial amplitude of the
systematic and random differences in Figs.~\ref{sdss_prince}--\ref{sdss_ciii}
led to the initiation of the investigation presented here.

\section{Master Quasar Template Construction}\label{sec:temcon}

The generation of the high-SNR quasar template to be used to calculate
cross-correlation redshifts begins with a sample of quasars at low
redshifts that possess emission line-determined redshifts. A somewhat
more involved procedure is then necessary to incorporate additional quasars
at higher redshifts into the master template. In this section the 
recipe for each element of the master template construction are outlined.

\subsection{Initial low-redshift quasar template}\label{sec:elin_start}

The narrow forbidden emission lines of \oiii $\lambda\lambda$4960,5008 are
prominent in many quasar spectra with redshifts $z$$<$0.8 and a composite
spectrum based on the combination of quasars with redshifts determined via the
location of \oiii emission forms the starting point for the construction of the
master quasar template.  The $\simeq$19\,000 quasars with SDSS redshifts
$z$$<$0.85 are searched for the presence of \oiii emission in a narrow
wavelength interval ($\simeq$100\,\AA) corresponding to the predicted rest-frame
location calculated from the SDSS redshift.

A `continuum' is defined using a median filter of 21 pixels and \oiii
$\lambda\lambda$4960,5008 emission then identified using a matched-filter
detection scheme applied to a continuum-subtracted version of each spectrum
\citep[e.g.][]{1985MNRAS.213..971H}.  \oiii emission is often broad and
frequently exhibits strong asymmetries \citep{1981ApJ...247..403H},
the small filter-scale adopted for the
\oiii detection is chosen with the aim of isolating narrow, well-defined, peaks
that may be present.  The filter template consists of two Gaussian components of
the same width, centred at 4960.30\,\AA \ and 5008.24\,\AA, with flux ratio 1:3.

Emission features can be identified reliably via detections with relatively low
SNR, particularly given the restricted wavelength range searched in each
spectrum.  However, given the importance of establishing accurate redshifts as
the first step in the construction of the composite quasar, the 8542
spectra possessing \oiii detections with SNR$\ge$8$\sigma$ form the starting
point for the template construction.

The recipe used to combine spectra with specified redshifts into a composite is
as follows:

\begin{itemize}
\item pixels falling within 6.0\,\AA \ of the strong night-sky lines at
5578.5\,\AA \ and 6301.7\,\AA \ are flagged
\item pixels without valid SDSS data, determined from the SDSS noise array
provided for each spectrum, are flagged
\item spectra are shifted to the rest-frame, with the native 
69\kms `pixels' of the original SDSS spectra retained. The signal from
each spectrum
is placed onto the master rest-frame wavelength array using a
`nearest pixel'-scheme, thereby avoiding the need for any rebinning or 
interpolation\footnote{The additional `jitter' that the simple nearest-pixel
scheme introduces is small, with a maximum error of 34.5\kms and an
increased dispersion of $\sigma$=20\kms, less than a third of a pixel, in
the extent of features in the resulting composite spectra.}
\item spectra are normalised using a wavelength interval common to all spectra
\item spectra are median-filtered with a window of 61-pixels to define a 
`continuum'. Spectrum pixels falling more than 4.5$\sigma$ below the continuum,
along with a grow-radius of two pixels, are flagged, effectively removing
wavelengths affected by strong narrow absorption
\item the median value of all non-flagged pixels at each rest-frame 
wavelength is calculated (a minimum of 100 spectra must contribute)
\end{itemize}

At this point a very high SNR composite quasar spectrum is available, extending
down to rest-frame wavelength $\lambda$$\simeq$2300\,\AA.  The \oiii emission
moves beyond the red limit of the SDSS spectra at $z$$>$0.8 and it is necessary
to use a cross-correlation scheme, employing a much greater wavelength range of
the quasar spectrum, to allow the construction of the master template further
into the rest-frame ultra-violet.

\subsection{Cross-correlation redshift algorithm}\label{sec:cc_calc}

The cross-correlation algorithm is based on a straightforward spatial
cross-correlation between an individual quasar spectrum and a high-SNR template
spectrum.  The key elements of the cross-correlation calculation are i) a
conservative choice of the portions of the quasar spectrum to employ in the
calculation, avoiding strong emission lines close to the edges of the observed
spectrum, and ii) application of an essentially identical `window' to both the
individual quasar and template spectra prior to the cross-correlation
calculation.

For each quasar spectrum, with its companion error-array, pixels are excluded
from the cross-correlation calculation according to a sequence of rules/tests.
The first and last valid pixels, where the SDSS spectrum error-array is not set
to 0, define the limits of the accessible wavelength range.  In the
observed-frame:

\begin{itemize}
\item the first 25-pixels at each end are excluded

\item pixels within 6\,\AA \ of each of the strong night-sky
emission lines at 5578.5\,\AA \ and 6301.5\,\AA \ are excluded

\item narrow absorption features are identified by examining a
continuum-subtracted spectrum. The continuum is defined using a
61-pixel median filter. Pixels that fall more than 4.5$\sigma$ below
the continuum are flagged and a grow-radius of 2-pixels then
applied. Thus, a single pixel exceeding the threshold results in the
exclusion of 5-pixels.
\end{itemize}

The quasar spectrum is then transformed to the rest-frame using the
specified redshift estimate, $z_{init}$\footnote{The SDSS-derived redshift
is used to determine the value of $z_{init}$ initially but all the
cross-correlation estimates are recalculated using an updated 
value of $z_{init}$ from the cross-correlation calculation 
itself. The cross-correlation estimates converge to the 10$^{-5}$ level
with just one iteration.}. In the rest-frame:

\begin{itemize}
\item pixels with $\lambda$$>$7000\,\AA \ are excluded

\item for $z_{init}$$>$0.38, pixels with $\lambda$$>$6400\,\AA,
i.e. the H$\alpha$ region, are excluded

\item for $z_{init}$$<$0.45, pixels with $\lambda$$<$2900\,\AA,
i.e. the \mgii region, are excluded

\item for $z_{init}$$<$1.10, pixels with $\lambda$$<$1975\,\AA,
i.e. the \ciii region, are excluded

\item for $z_{init}$$<$4.00, pixels with $\lambda$$<$1675\,\AA,
i.e. the \civ region, are excluded

\item pixels with $\lambda$$<$1275\,\AA, i.e. the \nv and
Lyman-$\alpha$ lines and the Lyman-$\alpha$ forest, are always
excluded.
\end{itemize}

Following the definition of the restricted wavelength interval over
which the quasar spectrum is retained, continua, estimated using a
large-scale, 601-pixel, median filter are subtracted from the quasar
and the template spectra. Exactly the same wavelength interval is used
to estimate the continua subtracted from the quasar and the template
spectra.

With continuum-subtracted quasar, $Q_{cs}$, and template, $T_{cs}$,
spectra available, the cross-correlation, for lag "l", is performed:

\be
cc(l)=
\frac{\sum_i Q_i T_i/\sigma_i^2}{\sqrt{{\sum_i (Q_i/\sigma_i)^2} {\sum_i (T_i/\sigma_i)^2}}}
\ee

\noindent where $\sigma_i$ is the noise, as provided in the SDSS
FITS-files, and the `$cs$' sub-scripts have been omitted for
clarity.

A quadratic fit is then made to the array of $cc(l)$ values over the interval
$l$=$\pm npix$, with $npix$=100.  The fit is then refined, performing quadratic
fits to narrower pixel intervals centred on the peak of the previous quadratic
fit, with the final fit determined over an interval of $l$=$\pm npix$/5.  The
output consists of a redshift estimate, $z_{fin}$, and a cross-correlation
amplitude, $cc_{max}$, in the range --1$\le$$cc_{max}$$\le$1, which
parameterises the degree of similarity between the two spectra.  Extensive
experimentation demonstrates that the requirement $cc_{max}$$\ge$0.2 results in
an almost error-free catalogue of cross-correlation redshifts.  However, it must
be stressed that the use of such a low amplitude is only possible because of the
extremely low occurrence of catastrophic redshift mis-identifications resulting
from the SDSS spectroscopic pipeline and the subsequent refinements of
\citet{2007AJ....134..102S, 2010AJ....XXX..XXXX}.

\subsection{Quasar template extension for redshifts 0.8$<$$z$$\le$1.6}

With the cross-correlation redshift determination procedure in place it is
possible to utilise quasars with redshift $z$$>$0.8 to extend the master
template.  To ensure that the template construction is not adversely affected by
the inclusion of spectra with poor SNR, or the presence of broad absorption line
(BAL) troughs, the full quasar sample (Section \ref{sec:samp}) was restricted to
those objects satisfying the following criteria:

\begin{itemize}
\item SDSS spectrum spectroscopic SNR, {\tt SN\_R}+{\tt SN\_I} $\ge$18.0
\item quasar not identified as BAL-quasars by 
\citet{2009ApJ...692..758G} or from our own BAL catalogue (Allen et al. 2010,
in preparation)
\end{itemize}

Application of the criteria reduce the sample by approximately a half, to
$\simeq$44\,500 spectra.  Cross-correlation redshifts are then calculated for
4071 spectra with 0.8$<$$z$$\le$1.0 according to the prescription of
Section~\ref{sec:cc_calc}.  All spectra with $cc_{max}$$\ge$0.2 and redshifts,
0.8$<$$z$$\le$1.0, are combined to produce a composite.  Then, the original and
new composites are combined by taking the average, weighted by the relative
number of spectra contributing at each wavelength.  The effect is
to determine redshifts for quasars using only the wavelength range where the
initial (lower-redshift) composite is of high SNR.  The procedure is then
repeated for intervals of $\Delta z$=0.2 up to redshift $z$=1.6.
Table~\ref{tab:cc_construct} summarises the number of spectra, median absolute
magnitudes and wavelength coverage for all of the composites used to generate
the final master template spectrum.

\begin{figure*}
\begin{minipage}{14cm}
 \includegraphics[scale=0.5]{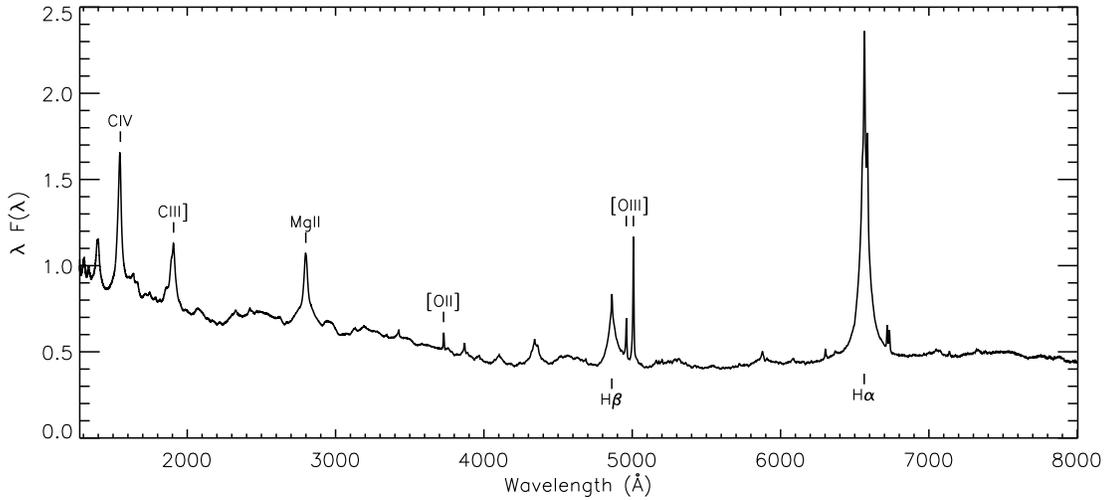}
\end{minipage}

\caption{The master quasar template, plotted as rest-wavelength {\it versus}
 $\lambda$F($\lambda$) (in arbitrary units).  The prominent emission line
 features of H$\alpha$, \oiii, H$\beta$, \oii, \mgii, \ciii + \siiii + \aliii
 and \civ are indicated.  }

\label{fig:qtemp_a}
\end{figure*}

The key elements of the scheme are the use of wavelength regions $\ge$1975\,\AA
\ for the calculation of redshifts of quasars up to $z$=1.6, thereby excluding
the \ciii and \civ emission lines. The rest-frame
ultra-violet region of interest is shown in Fig.~\ref{fig:qtemp_a}.

\subsection{Luminosity-dependent emission line shifts}\label{sec:em_shifts}

\begin{figure}
 \begin{minipage}{\textwidth}
 \includegraphics[scale=0.30,angle=-90]{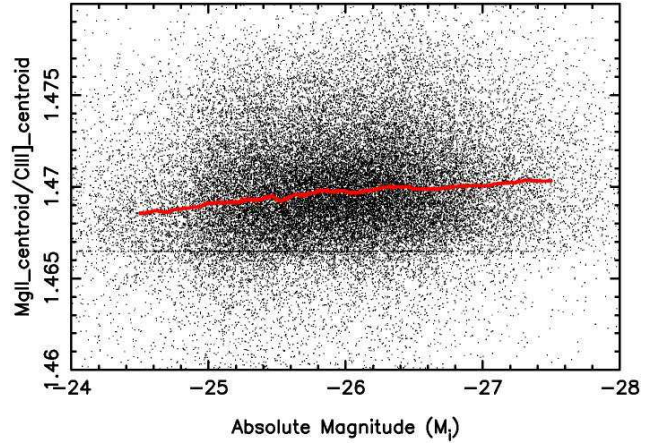}
  \end{minipage}
  \caption{The observed-frame ratio of the centroid wavelengths for \mgii
  and \ciii emission lines as a function of the quasar absolute magnitude, $M_i$.
  The quasar sample consists of 42\,609 objects with redshifts
  1.1$<$$z$$<$2.2. The solid red line,
calculated using a 2001-point running median of the data-points, shows
the strong systematic trend with quasar luminosity.  Data for 1000 spectra lie 
outside the x-y range plotted.}
\label{fig:mg_ciii}
\end{figure}

Quasar luminosity-dependent systematic effects related to the rest-frame
locations of \caii absorption, \oii, \oiii and \mgii emission are at or below
the level of 30\kms (Section~\ref{sec:systematics}).  However, the same is not
true when considering the location of the \mgii emission line and the next
prominent emission line complex of \ciii $\lambda$1908, \siiii $\lambda$1892 and
\aliii $\lambda$1857 as one moves further into the ultra-violet.
Fig.~\ref{fig:mg_ciii} shows the ratio of the observed-frame
centroids\footnote{The line centroids are generated as part of the SDSS {\tt
spectro1d} pipeline.}  of `\ciii' and \mgii for quasars in the redshift interval
1.1$<$$z$$<$2.2, where both lines are present in the SDSS spectra.  The
systematic trend of $\simeq$2$\times$10$^{-3}$ in the wavelength ratio as a
function of quasar luminosity translates directly into a systematic in $\Delta
z$/(1+$z$) where the \ciii emission line region contributes significantly to the
cross-correlation signal.

The luminosity-dependent change in the relative positions of the \mgii and \ciii
emission lines presents a problem when considering the generation of the master
quasar template.  Indeed, the large systematic means that care is needed when
calculating redshifts for a population of quasars with a range of luminosities.
Consider the result of cross-correlating a quasar template with a quasar of high
luminosity.  The \ciii emission in the quasar is at slightly smaller rest-frame
wavelength than in the template and the resulting redshift estimate will lie
somewhere between redshift estimates based on the locations of the \mgii and
\ciii lines alone.  Performing such cross-correlation redshift estimates for a
sample of higher redshift quasars and then updating the template with their
spectra will have the effect of biasing the profile/location of the \mgii
emission line to smaller wavelengths.  The effect is pernicious in that
subsequent use of such a template to calculate redshifts for quasars where \ciii
is not even visible, will produce biased values because of the systematic change
in the profile/location of \mgii emission and other features in the template.
Similar, even larger, luminosity-dependent systematic trends are also present in
the relative locations of the \ciii emission complex and the \civ emission line.

The existence of the systematic luminosity-dependent trends in the location of
the \ciii and \civ lines means that care must be taken in the definition of the
master quasar template. 

\subsection{Quasar template extension for redshifts 1.6$<$$z$$\le$2.0}

As cross-correlation redshifts are calculated using only rest-frame wavelengths
$>$1975\,\AA \ for quasars with redshifts $z$$<$1.6, the current master template
is free of the luminosity dependent systematics described above.  For
redshifts $z$$\ge$1.6 it is necessary to use the wavelength range including the
\ciii emission complex to increase the SNR of the cross-correlation signal.
However, to preserve the form of the composite at wavelengths $>$2000\,\AA,
quasars with redshifts 1.6$\le$$z$$<$2.0 are incorporated into the template
using only wavelengths $<$2000\,\AA.  The master quasar template is extended
down to $\lambda$=1275\,\AA \ in two redshift slices
(Table~\ref{tab:cc_construct}), with the stricture that the quasars in the
redshift interval 1.6$\le$$z$$<$2.0 are not allowed to contribute to the
template at wavelengths $\ge$2000\,\AA.

The result is a final master template in which the form of the spectrum at
$\lambda$$\le$2000\,\AA \ is appropriate to a quasar of a particular absolute
magnitude ($M_i$$\simeq$--26)\footnote{Absolute magnitudes, $M_i$, are calculated
using the prescription of \citet{2007AJ....134..102S}.}.  Given that the SDSS quasars possess a
significant spread in absolute magnitude, systematic trends in the redshift
determinations using the master quasar template are expected but the amplitude
and form of the systematic trends are such that reliable corrections are
possible (Section~\ref{sec:cc_newzs}) and reliable redshifts can be derived
for objects with redshifts up to $z$=4.5.

\begin{table*}
\begin{center}
\caption{\label{tab:cc_construct} Quasar cross-correlation template 
definition parameters}

\begin{tabular}{ccccccc} \hline
Redshift Range & Median $M_i$ & Number & FD Number & Wavelength Coverage & Redshift Method & Wavelength Contribution \\
               &              &        &           & (\AA)               &                 & (\AA)                    \\
\hline
0.0--0.4 & -22.50 & 3958 & 3958 & 2732--8004 & \oiii & 2732--8004 \\
0.4--0.8 & -23.86 & 4584 & 4584 & 2136--6550 & \oiii & 2136--6550 \\
0.8--1.0 & -24.89 & 4071 & 4071 & 1908--5099 & \mgiicc ($>$1975\AA) & 1908--5099 \\
1.0--1.2 & -25.38 & 4762 & 393 & 1732--4590 & \mgiicc ($>$1975\AA) & 1732--4590 \\
1.2--1.4 & -25.77 & 5118 & 375 & 1589--4176 & \mgiicc ($>$1975\AA) & 1589--4176 \\
1.4--1.6 & -26.15 & 5087 & 341 & 1466--3819 & \mgiicc ($>$1975\AA) & 1466--3819 \\
1.6--1.8 & -26.45 & 3882 & 262 & 1363--3534 & \mgiicc ($>$1675\AA) & 1363--2000 \\
1.8--2.0 & -26.71 & 2797 & 169 & 1275--3275 & \mgiicc ($>$1675\AA) & 1275--2000 \\
\hline
\end{tabular}
\vspace*{-0.4cm}
\end{center}
\end{table*}

\section{New Quasar Redshifts}\label{sec:cc_newzs}

Redshift determinations, in decreasing order of accuracy and increasing quasar
redshift, can be obtained using \oiii emission lines, \oii emission lines,
cross-correlation including the \mgii emission line (\mgiicc), cross-correlation
including the \ciii emission line complex (\ciiicc) and, finally,
cross-correlation including the \civ emission line (\civcc).  For the
cross-correlation results, empirical comparisons of redshifts derived using
different rest-frame wavelength regions allow conversion relations to be derived
as a function of quasar absolute magnitude and redshift.  The goal is to build a
redshift `ladder' for quasars of increasing redshift that allows the redshift
estimates to be placed on the same underlying systemic reference system.  The
sub-Sections below consider in turn each step in the ladder.

\subsection{\oii and \oiii narrow emission line redshifts}

Redshifts for 13291 quasars with redshifts $z$$<$0.84 are available via the
detection of \oiii $\lambda\lambda$4960,5008 emission with a SNR$\ge$6.0$\sigma$
(Section~\ref{sec:elin_start}).  Similarly, detection of the \oii
$\lambda\lambda$3727,3729 emission doublet at a SNR$>$6.0$\sigma$ provides
redshift determinations for an additional 3844 quasars, with redshifts
$z$$<$1.31.

In the case of \oii detections, a single Gaussian, centred at 3728.60\,\AA, is
used.  The \oii doublet consists of two components centred at 3727.09\,\AA \ and
3729.88\,\AA.  The observed component ratio varies from quasar to quasar but is
normally in the range 0.8:1--0.9:1, leading to the effective wavelength of
3728.60\,\AA \ adopted.

\subsection{Cross-correlation redshifts including \mgii}

\mgiicc redshifts are available for a further 12289 quasars with
redshifts $z$$\le$1.10. The minimum rest-frame wavelength involved in the
cross-correlation is 1975\,\AA \ and systematic offsets relative to the 
emission line redshifts generated in the
previous sub-Section are not predicted or detectable.
 
In the interval 1.1$<$$z$$<$2.1, the \mgiicc redshifts involve rest-frame
wavelengths below 1800\,\AA \ and the signal is increasingly dominated by the
\ciii emission as redshift increases and the \mgii line shifts into the far red
of the SDSS spectra.  Additionally, there is also the luminosity-dependent
variation in the location of the \ciii emission complex to take into account.
The amplitude of the systematic redshift bias is small, only 
$\simeq$2$\times$10$^{-4}$ too large at the highest redshift $z$=2.1.

Taking the \mgiicc redshifts and the corresponding \mgii emission line centroid
determinations from the SDSS pipeline allows the dependence of the
cross-correlation bias on redshift and absolute magnitude to be quantified.
Treating the systematic as separable in redshift and luminosity, a sample of
$\simeq$42\,000 quasars, in the redshift interval 1.1$<$$z$$<$2.1, shows that
linear corrections to the raw cross-correlation redshifts, with slopes of
1.61$\times$10$^{-4}$ per unit redshift and 7.2$\times$10$^{-5}$ per unit
magnitude (reducing the raw redshift estimates as redshift and luminosity
increase), bring any residual systematic trends down to the
$\simeq$1$\times$10$^{-5}$ level.  Note that the sense and amplitude of the
difference between the raw \mgiicc redshifts and the \mgii emission line
centroids are entirely consistent with the existence of the luminosity-dependent
emission line shifts and the way that the master quasar template spectrum is
constructed (Section~\ref{sec:em_shifts}).  Corrected \mgiicc redshifts are
available for 43\,728 quasars in the interval 1.1$<$$z$$\le$2.1.

\begin{figure}
 \begin{minipage}{\textwidth}
  \includegraphics[scale=0.30,angle=-90]{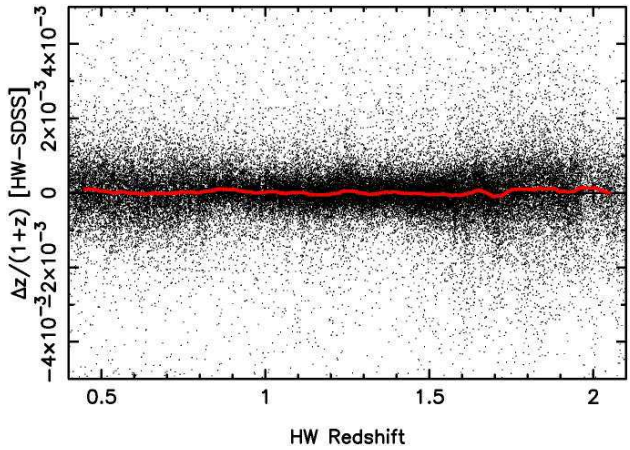}
  \end{minipage}
  \caption{Redshift differences,$\Delta z/(1+z)$, between HW
   redshifts and
redshifts derived from the SDSS-determined \mgii emission line locations.
Spectra plotted possess \mgii emission line SNR$>$10.  The solid red line,
calculated using a 2001-point running median of the data-points, demonstrates
the complete removal of detectable systematic trends with redshift.   Contrast the behaviour
with that shown in Fig.~\ref{sdss_mgii}.
Data for 60\,120 spectra are
included (559 lie outside the y-axis range plotted).}
\label{hw_mgii}
\end{figure}

\subsection{Cross-correlation redshifts including \ciii}

At redshifts $z$$>$2.1 the \mgii line no longer contributes to the
cross-correlation redshifts and the full effect of the systematic variation in
the rest-frame locations of the \mgii and \ciii emission lines must be taken
into account.  Fortunately, an empirical determination of the systematic
differences between the corrected, un-biased, redshifts derived above and
cross-correlation redshifts using only the rest-wavelength region
1675$<$$\lambda$$<$2650 \AA, termed \ciiicc redshifts, is straightforward to make.

\begin{figure}
 \begin{minipage}{\textwidth}
 \includegraphics[scale=0.30,angle=-90]{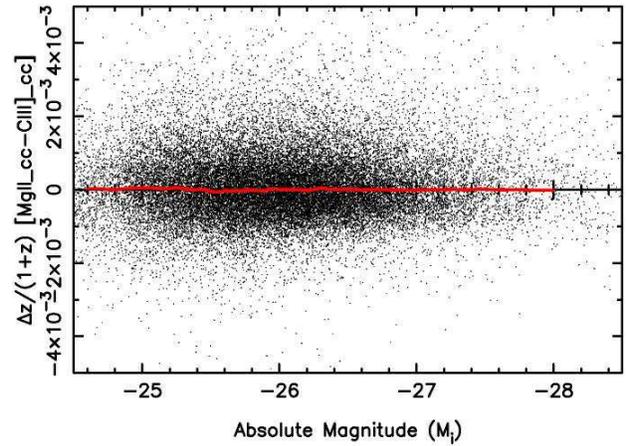}
  \end{minipage}
 \caption{Redshift differences,$\Delta z/(1+z)$, between corrected \mgiicc
redshifts and corrected \ciiicc redshifts derived using the \ciii emission line
complex alone, as a function of absolute magnitude.  The solid red line is a
2001-point running median of the data-points, illustrating the absence of any
detectable systematic trends as a function of quasar absolute magnitude.  Data
for 34\,891 spectra are included (119 lie outside the y-axis range plotted).}
\label{hw_ciii_mabs}
\end{figure}

\begin{figure}
 \begin{minipage}{\textwidth}
  \includegraphics[scale=0.30,angle=-90]{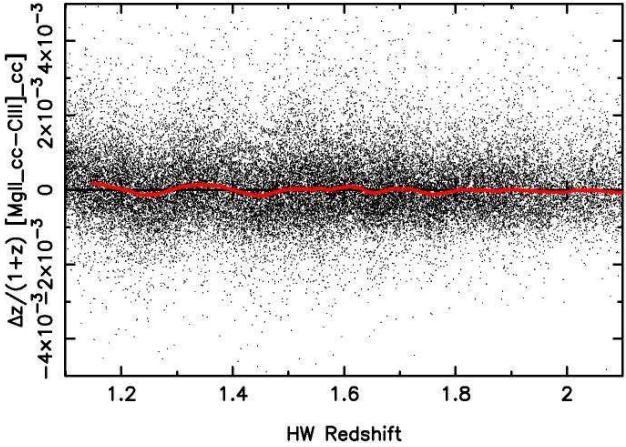}
  \end{minipage}
 \caption{Redshift differences,$\Delta z/(1+z)$, between the corrected \mgiicc
redshifts and corrected \ciiicc redshifts derived, as a function of redshift.
The solid red line is a 2001-point running median of the data-points showing the
absence of any detectable systematic trends as a function of quasar redshift.
Data for 35\,179 spectra are included (120 lie outside the y-axis range
plotted).}
\label{hw_ciii_z}
\end{figure}

The differences between the corrected \mgiicc redshifts and raw \ciiicc
redshifts, derived using a maximum rest-frame wavelength of $\lambda$=2650\,\AA,
i.e.  excluding the \mgii emission line, are available for $\simeq$35\,000
quasars with 1.1$<$$z$$<$2.0.  The difference as a function of quasar absolute
magnitude is systematic and well represented by a linear trend; a linear fit has
slope 1.67$\times$10$^{-4}$, increasing the raw redshifts for increasing
bright quasars.  Application of the correction removes any detectable systematic
effects as a function of absolute magnitude (Fig.~\ref{hw_ciii_mabs}) or
redshift (Fig.~\ref{hw_ciii_z}).  The same correction is then applied to \ciiicc
redshifts for 13859 quasars with $z$$>$2.1.

\subsection{Cross-correlation redshifts including \civ}\label{sec:civ_corr}

The intention throughout is to avoid the use of the rest-frame wavelength region
including the \civ emission line, which is known to show large asymmetric
variations in shape, and hence of the line centroid.  However, for 3274 quasars
in the redshift interval 2.1$<$$z$$<$4.5, the cross-correlation signal from the
rest-frame $\lambda$$>$1675\,\AA \ region is too low to produce a reliable
\ciiicc redshift. For these 3274 quasars a cross-correlation redshift determination 
employing the rest-frame wavelength interval $\lambda$$>$1275\,\AA \ 
(\civcc) is made.

There is a strong luminosity-dependent bias present due to the systematic
variation in the location of the rest-frame \civ emission line centroid.  The
situation is complicated by the presence, in a large number of quasars, of
significant absorption blueward of the \civ emission centroid, which biases the
line centroid to the red.  To decouple the effects on the \civ emission line of
quasar luminosity and the presence of absorption, a sub-sample of $\simeq$25\,000
quasars with essentially undetectable absorption blueward of the \civ emission
line centroid is defined\footnote{The absorption strength is parameterised using
an integrated absorption equivalent width (AEW), calculated over the velocity
range -29\,000 to 0\kms, relative to the predicted \civ $\lambda$1549 location}.

\begin{figure}
 \begin{minipage}{\textwidth}
 \includegraphics[scale=0.30,angle=-90]{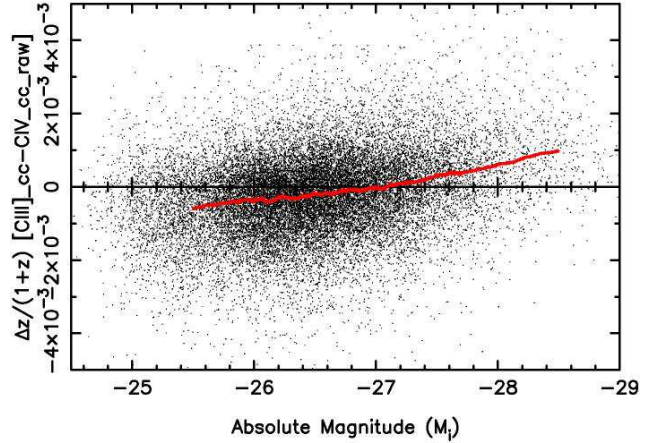}
  \end{minipage}
 \caption{Redshift differences,$\Delta z/(1+z)$, between the corrected HW
 cross-correlation redshifts and uncorrected \civcc redshifts derived
 including the \civ emission line, as a function of absolute magnitude.
   The solid red line is a 2001-point running median of the data-points.  
   Data for 34\,956 spectra are included (119 lie outside the y-axis range 
   plotted).
 }
\label{ciii_civ_raw}
\end{figure}

An empirical correction for the systematic luminosity-dependent redshift bias
(Fig~\ref{ciii_civ_raw}) can then be made in exactly the same way that the
\ciiicc redshifts were referenced to the un-biased system.  A two-part linear
fit to the absolute magnitude dependence, with slope 6.67$\times$10$^{-4}$
for $M_i$$<$--27.0 and slope 3.90$\times$10$^{-4}$ for $M_i$$\ge$--27.0,
provides an excellent fit to the systematic trend.  The raw \civcc redshift
estimates are increased for more luminous quasars, reducing systematic redshift
differences to undetectable levels.  The amplitude of the systematic bias in the
raw \civcc redshifts is large, a factor of four greater than the \ciii
dependence, and the correction results in a substantial reduction in the
dispersion between the \civcc and un-biased redshift determinations.

Objects with significant absorption blueward of the \civ emission line, many of
which are Broad Absorption Line (BAL) quasars, show an extended tail of redshift
deviations to high values.  A second systematic correction, as a function of the
absorption equivalent width (AEW) is then made, using the differences between
the corrected \ciiicc redshifts and the absolute magnitude corrected \civcc
values for $\simeq$38\,000 quasars.  The actual correction applied is based on
the empirically-determined median $\Delta z/(1+z)$ {\it versus} AEW relation but
the amplitude of the well-determined correction is closely reproduced by a
linear fit with slope --2.5$\times$10$^{-5}$ over the 0--200 range of AEW used.
The additional quasar-to-quasar dispersion in redshift at large AEW is
significant but only 574 quasars with AEW$>$20 in the final catalogue possess 
\civcc redshifts.

Table~\ref{tab:z_corr} summarises the redshift and absolute magnitude dependent 
corrections made to the redshifts from the different estimation schemes in the 
ladder. 

\begin{table*}
\begin{center}
\caption{\label{tab:z_corr} Systematic redshift corrections}

\begin{tabular}{ccccc} \hline
Redshift Method & Number & Redshift Interval & Redshift Correction& $M_i$ Correction \\
                &        &                   & ($|\Delta z/(1+z)|$ per unit $z$)       & ($|\Delta z/(1+z)|$ per unit $M_i$)    \\
\hline

\oiii    & 13291 & $<$0.84  & --                   & --                   \\
\oii     &  3844 & $<$1.35  & --                   & --                    \\
\mgiicc & 12289 & $<$1.1    & --                   & --                  \\
\mgiicc & 43728 & 1.1--2.1  & 1.61$\times$10$^{-4}$ & 7.2$\times$10$^{-5}$ \\
\ciiicc & 13859 & 1.1--4.1  & --                      & 1.67$\times$10$^{-4}$ \\
\civcc  & 3274 & 1.5--5.5   & --                      & 6.67$\times$10$^{-4}$ $<$--27.0 \\
\civcc  &      &            & --                      & 3.90$\times$10$^{-4}$ $\ge$--27.0\\
\hline
\end{tabular}
\vspace*{-0.4cm}
\end{center}
\end{table*}

\subsection{Additional redshifts}

Redshifts for an additional 124 quasars are available, although one of
the strict criterion described above is not satisfied, e.g. SNR$<$6.0$\sigma$
for an emission line detection, or $cc_{max}$$<$0.2. These redshifts are
included in the catalogue but are highlighted by the inclusion of a special
status flag.

Finally, the emission line detection and cross-correlation schemes fail to
provide reliable redshift estimates for 1256 quasars.  These objects consist
primarily of a mix of pathological spectra, including extreme BAL quasars, and
spectra of very low SNR.  The objects are included in the redshift catalogue for
completeness, with redshifts and redshift errors taken from the Schneider et al.
catalogues \citep{2007AJ....134..102S, 2010AJ....XXX..XXXX} and the {\tt
spectro1d} pipeline for DR6.  Again, the source of the redshifts is indicated in
the catalogue via a status flag.

\section{Systemic Redshifts and Redshift Uncertainties}\label{sec:systematics}

Section~\ref{sec:cc_newzs} describes the scheme adopted to obtain redshifts
using the most reliable estimation procedure for each quasar.  Redshifts are
referenced to the zero-point provided by the location of the
\oiii~$\lambda\lambda$4960,5008 emission lines.  The goal is to reduce
systematic errors in $\Delta z/(1+z)$ to the level of $\le$1$\times$10$^{-4}$
(30\kms) per unit redshift and $\le$2.5$\times$10$^{-5}$ (8\kms) per unit
absolute magnitude.  In this section the question of referencing the \oiii
emission line redshifts to the systemic system defined by the quasar host
galaxies is considered.  Starting with the comparison of absorption line and
\oiii emission redshifts, the uncertainties in redshift estimates arising from
both the intrinsic quasar-to-quasar variation and the reproducibility of the
determinations for each technique in the ladder are quantified.

\subsection{\oiii and host galaxy systemic redshifts}\label{sec:caii_host}

Redshift estimates based on the detection of photospheric absorption
from stars in the spatially averaged spectrum of the quasar host
galaxy might be expected to provide a close to `ideal' systemic
redshift.  Given the nature of the SDSS spectra, coupled with the
large luminosity of the quasars at rest-frame optical and near-ultra-violet
wavelengths, detection of photospheric absorption is not possible for
the majority of objects.  However, a direct comparison of redshifts
derived from photospheric \caii $\lambda\lambda$3934.8,3969.6 and
\oiii emission is possible for a sample of objects with redshifts
$z$$<$0.4.  Generating a catalogue of \caii absorption detections with
SNR$>$6$\sigma$, matched to quasars with \oiii detections in the
redshift interval 0.2$\le$$z$$\le$0.4, produces 825 quasars.
Restricting the sample to objects that satisfy the
luminosity-criterion for inclusion in the Schneider et al.  SDSS
quasar compilations, results in 615 quasars with absolute magnitudes
covering the full range --24.0$<$$M_i$$<$--22.0.  Composite spectra with
median $M_i$=--22.2 and --22.6 possess both \caii absorption and \oiii
emission at high SNR.

Measuring the centroid of the strong \caii K-line at 3934.8\AA \ (\caii H is
blended with H$\epsilon$ absorption, producing a shift to longer
wavelengths) and the centroid of the \oiii $\lambda\lambda$4960.30,5008.24
emission, measured above the 50 per cent peak-height level, shows that the
\oiii emission is shifted by 45$\pm$5\kms to the blue, with no detectable
dependence on luminosity.  The offset determined from the composite spectra
is in good agreement with the results of \citep{2005AJ....130..381B} and the
distribution of individual \caii and \oiii redshift-differences for the 615
objects contributing to the composites.  The median redshift difference for
the sample corresponds to a velocity shift of 38$\pm$9\kms blueward for
\oiii relative to \caii.  We therefore correct all redshifts by a 45\kms
shift to the red to bring the zero-point into coincidence with the system
defined by the \caii K-line.

After correcting for the contribution of the redshift reproducibility
($\sigma_{repro}$=2.1$\times$10$^{-4}$, calculated using multiple spectra
(Section~\ref{sec:samp})) in estimating the \caii redshifts, the empirically
determined quasar-to-quasar scatter\footnote{All root-mean-square (rms), or
$\sigma$, values are calculated using absolute differences,
$|x_i$--median($x$)$|$, with an iterative rejection scheme that removes values
$>$4$\sigma$, up to a maximum of three iterations.  In fact, the parameter
distributions are not in general significantly non-Gaussian, with a maximum of
$\simeq$5 per cent of values excluded from the final rms-estimates} of \oiii
redshifts about the \caii absorber redshifts is
$\sigma_{intrin}$=3.5$\times$10$^{-4}$.

\subsection{\oiii and \oii emission line redshifts}\label{sec:oxcomp}

A similar comparison can be made between the \oiii and \oii emission line
derived redshifts for the more than 7500 quasars, redshifts $z$$<$0.8, with
both \oiii and \oii emission line detections.  A small systematic velocity
offset is present, with the \oii derived redshifts 24$\pm5$\kms redward of
the \oiii derived redshifts.  Thus, relative to the systemic reference
defined by the \caii K absorption, the offsets are 21$\pm$5\kms blueward for
\oii and 45$\pm$5\kms blueward for \oiii, in excellent agreement with
previous work \citep[e.g.]{2005AJ....130..381B}.

Comparison of 1378 (1103) spectrum pairs results in median errors of
5$\times$10$^{-5}$ and 1.4$\times$10$^{-4}$ in the reproducibility of $\Delta
z$/(1+$z$) for \oiii and \oii respectively.  The smaller error for \oiii results
from the typically higher SNR of the emission compared to the \oii line.

Systematic trends, derived from linear fits to the redshift differences between
\oiii and \oii redshifts, are $\Delta z$/(1+$z$)=2.1$\times$10$^{-5}$ per
magnitude and $\Delta z$/(1+$z$)=5.9$\times$10$^{-5}$ per unit redshift.  The
sense is that redshifts derived from the \oiii emission lines become
systematically smaller for objects with higher luminosities(redshifts).  The
luminosity-dependent systematic is stronger than for redshift given the dynamic
ranges of the variables present in the sample.  The trend is consistent with an
increasing degree of blue-asymmetry present in the \oiii emission lines at
increasing quasar luminosity.  The effect is, however, small, amounting to a
maximum of $\pm$4.5$\times$10$^{-5}$, or, $\pm$14\kms, about the mean relation
for the sample.

The empirically determined quasar-to-quasar rms-scatter of the individual \oii
redshifts about the \oiii redshifts is $\sigma_{intrin}$=1.0$\times$10$^{-4}$
(30\kms).

\subsection{\oiii and \mgiicc redshifts}

\oiii emission line redshifts compared to \mgiicc redshifts, calculated {\it
excluding} the \oiii emission lines from the quasar template, for
$\simeq$12\,500 quasars with \oiii emission line reshifts shows an undetectable
offset, median $|\Delta z$/(1+$z$)$|$$<$10$^{-5}$.  Systematic trends, from
linear fits to the redshift differences, in sense \oiii - \mgiicc redshifts, are
$\Delta z$/(1+$z$)=+1$\times$10$^{-5}$ per magnitude and $\Delta
z$/(1+$z$)=-1.1$\times$10$^{-4}$ per unit redshift.  Both luminosity and
redshift parameterisations lead to systematics of at most
$\pm$3$\times$10$^{-5}$ (10\kms), for the dynamic ranges present in the
sample, more than an order of magnitude below the uncertainties in the
individual \oiii redshifts (Section~\ref{sec:caii_host}).  The sense and
amplitude of the small systematic is consistent with the results from the \oiii
to \oii emission line comparison (Section~\ref{sec:oxcomp}) and is again almost
certainly due to the increasing degree of blue-asymmetry present in the \oiii
emission lines at increasing quasar luminosity.  The comparison shows that the
\mgiicc redshifts are tied to the reference \oiii emission line redshifts to
very high accuracy and that any systematics present are at most 10$^{-4}$ in
$\Delta z$/(1+$z$), or 30\kms in velocity.

The empirically determined quasar-to-quasar scatter of the individual \mgiicc
redshifts about the \oiii redshifts is $\sigma_{intrin}$=2.5$\times$10$^{-4}$ or
75\kms, which represents an improvement of a factor of $\sim$3 compared to
careful determinations of the \mgii emission line location
\citep[e.g.][]{2008MNRAS.386.2055N}.

\begin{table*}
\begin{center}
\caption{\label{tab:z_error} Redshift uncertainties}

\begin{tabular}{cccccc} \hline
Redshift Method & Number & Redshift Interval & Reproducibility & Population rms & Cumulative Population rms\\
                &        &                   & ($\Delta z/(1+z)$) & ($\Delta z/(1+z)$) & ($\Delta z/(1+z)$)    \\
\hline
\caii K   & --    & 0.2--0.4 & 2.1$\times$10$^{-4}$ &  --  & -- \\
\oiii    & 13291 & $<$0.84  & 5$\times$10$^{-5}$   & 3.5$\times$10$^{-4}$ & 3.5$\times$10$^{-4}$ \\
\oii     &  3844 & $<$1.35  & 1.4$\times$10$^{-4}$ & 1.0$\times$10$^{-4}$ & 3.65$\times$10$^{-4}$ \\
\mgiicc & 12289 & $<$1.1   & 1.4$\times$10$^{-4}$ & 2.5$\times$10$^{-4}$ & 4.3$\times$10$^{-4}$ \\
\mgiicc & 43728 & 1.1--2.1 & 3.6$\times$10$^{-4}$ & 2.5$\times$10$^{-4}$ & 4.3$\times$10$^{-4}$ \\
\ciiicc & 13859 & 1.1--4.1 & 6.5$\times$10$^{-4}$ & 2.5$\times$10$^{-4}$ & 4.3$\times$10$^{-4}$ \\
\civcc  & 2700 & 1.5--5.5  & 5.1$\times$10$^{-4}$ & 8.0$\times$10$^{-4}$ & 9.1$\times$10$^{-4}$ \\
\civcc + AEW   & 574   &           &               & 1.0$\times$10$^{-3}$ & 1.4$\times$10$^{-3}$ \\
extra\_cc   & 124  & 0.5--4.5 & 6.0$\times$10$^{-4}$ & 3.0$\times$10$^{-4}$ & 3.5$\times$10$^{-4}$ \\
SDSS       & 1256 & 0.3--5.5 & as per SDSS & -- \\
\hline
\end{tabular}
\vspace*{-0.4cm}
\end{center}
\end{table*}

\subsection{\mgiicc and \ciiicc redshifts}

The luminosity-dependent correction to bring the \mgiicc and \ciiicc 
redshifts into coincidence is highly successful, as evidenced by  
Fig.~\ref{hw_ciii_mabs} and Fig.~\ref{hw_ciii_z}.

After allowing for the uncertainty in the determination of the \mgiicc and
\ciiicc redshifts due to the limited SNR of the SDSS spectra there is no
evidence for an additional quasar-to-quasar redshift scatter associated with the
use of the \ciii emission line region alone.

\subsection{\ciiicc and \civcc redshifts}

\begin{figure}
 \begin{minipage}{\textwidth}
  \includegraphics[scale=0.30,angle=-90]{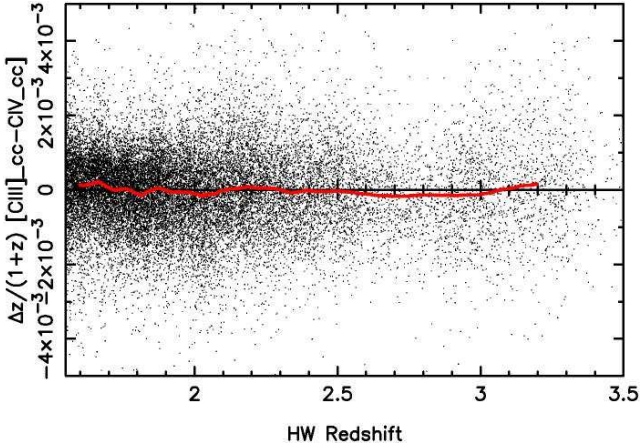}
  \end{minipage}
  
\caption{Redshift differences,$\Delta z/(1+z)$, between the corrected \ciiicc
redshifts and corrected \civcc redshifts derived using the \civ emission line complex
alone, as a function of redshift.  Data for 25\,008 quasars with AEW$\le$20 are
included (60 lie outside the y-axis range plotted).  The solid red line is a
2001-point running median of the data-points.  The offset between the two
redshift estimators is minimal except for the systematic `dip' centred at
$z$$\simeq$2.8, coincident with the significant drop in detection efficiency for
non-BAL quasars in the SDSS.}
\label{ciii_civ_z}
\end{figure}

The diversity of the form of the \civ emission line has been known through many
studies going back decades.  The removal of the systematic quasar
luminosity-dependent behaviour, amounting to $\simeq$650\kms (over four
magnitudes in quasar absolute magnitude), improves \civcc redshifts
considerably.  However, even for quasars with small AEW values the empirically
determined quasar-to-quasar scatter of the individual \civcc redshifts about the
\ciiicc redshifts is $\sigma_{intrin}$=8.0$\times$10$^{-4}$.  The situation is
much worse for quasars with significant absorption blueward of the \civ emission
line centroid.  The systematic correction applied reaches a full
5$\times$10$^{-3}$ for the most affected quasars and the dispersion at fixed AEW
value adds an additional rms-scatter of $\sigma_{intrin}$=1$\times$10$^{-3}$.
The number of quasars with large AEW values where only \civcc redshifts are
available is small, just 574 objects, but the associated redshift uncertainty is
accordingly large.

Fig.~\ref{ciii_civ_z} shows the redshift differences between the corrected
\ciiicc and corrected \civcc redshifts for 25\,008 quasars with only modest
absorption (AEW$\le$20) blueward of the \civ emission line.  The small amplitude
of the systematic differences illustrates the success of the statistical
correction to the initial \civcc redshifts.  However, the systematic negative
trend centred on $z$$\simeq$2.8, which reaches $\simeq$--1.8$\times$10$^{-4}$
(55\kms), illustrates the limitations of the correction.  The SDSS quasar
selection experiences a dramatic drop in efficiency over the interval
$z$$\simeq$2.7--2.9 that is much greater for non-BAL quasars than for BAL
quasars.  Thus, the overall number of quasars drops significantly, while the
BAL-fraction increases significantly.  There are only 13 quasars with
2.7$\le$$z$$\le$2.9 and with AEW$\le$20, in the final catalogue but a systematic
of at least --2$\times$10$^{-4}$ may well also be present among the 574 quasars
with \civcc redshifts and AEW$>$20

\begin{figure}
 \begin{minipage}{\textwidth}
  \includegraphics[scale=0.30,angle=-90]{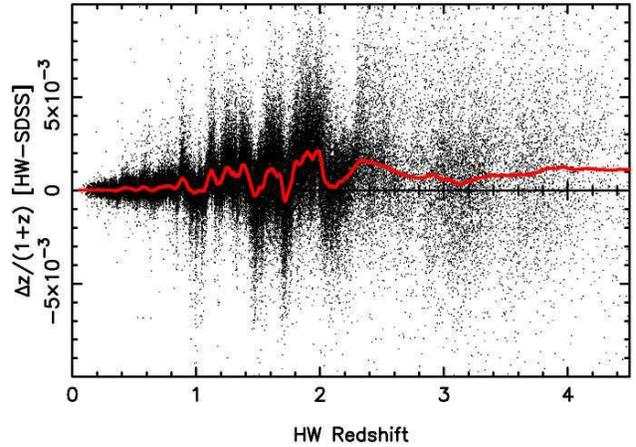}
  \end{minipage}
  
\caption{Redshift differences,$\Delta z/(1+z)$, between the final HW-redshifts
and SDSS-redshifts, as a function of redshift.  Note the large range on the
y-axis. Data for 90\,409 quasars are
included (1355 lie outside the y-axis range plotted).}
\label{fig:hw_sdss}
\end{figure}

\subsection{Summary}

All redshift estimates, except those taken direct from the SDSS, have been
increased by 45\kms (Section~\ref{sec:caii_host}) to bring the \oiii emission 
line based estimates onto the systemic system defined by the \caii K absorption.

Based on the large sample of repeat spectra, the internal reproducibility of
the new cross-correlation redshifts represent an improvement of more than a
factor 2 over the SDSS redshift values.  The reproducibility of the new
redshifts is indistinguishable from that for the Princeton redshift values
up to redshift $z$$\simeq$1.6.  At higher redshifts the Princeton algorithm
employs more information, via inclusion of the \civ emission line region at
$\lambda$$<$1675\,\AA, and results in significantly better reproducibility
for redshifts $z$$>$2.0 than in the scheme presented here.  However, the
large quasar-to-quasar variation contributing to the redshift uncertainty
(Table~\ref{tab:z_error}) means that differences in the internal
reproducibility are not a critical factor for the redshifts in the final 
catalogue.

Table~\ref{tab:z_error} summarises the uncertainties for the different
redshift estimates. Accurate redshift reproducibility estimates are available 
for all quasars, based on emission line SNR or $cc_{max}$ value, and are
incorporated in the redshift errors included in Table~\ref{tab:cat}. To
provide an indication of the relative contributions of the internal and
quasar-to-quasar uncertainties, Col. 4 of Table~\ref{tab:z_error} lists 
the median internal error for each redshift estimate. Col. 5
gives the quasar-to-quasar error. The quasar-to-quasar errors, working
from the chosen \caii absorption reference, have been
added in quadrature to produce the cumulative quasar-to-quasar 
uncertainties in Col. 7.

\section{Quasars with detections in FIRST}\label{radio_qsos}

The procedures described in Sections~\ref{sec:cc_newzs} and
\ref{sec:systematics} reduce systematic redshift errors as a function of
redshift and absolute magnitude by more than order of magnitude compared to the
publicly available SDSS redshifts.  However, a further significant reduction in
the remaining relatively large quasar-to-quasar uncertainties will require a
detailed investigation of the spectral energy distribution (SED) dependent
changes in the properties of the most prominent emission lines (which dominate
the redshift determinations).  Such an investigation is beyond the scope of this
paper but it is relatively straightforward to consider systematic redshift
differences that correlate with the detection of SDSS quasars in the Faint
Images of the Radio Sky at Twenty centimetres
\citep[FIRST,][]{1995ApJ...450..559B}.

For redshifts $z$$<$1.1, the systematic differences between the
populations of FIRST-detected (FD) and not-FIRST-detected (nFD) quasars with
\mgiicc cross-correlation redshifts are at the $\Delta z$/(1+$z$)$\le$10$^{-4}$
level.  However, for redshifts $z$$>$1.1, once the \ciii+\siiii+\aliii emission
line complex contributes to the redshift determination, systematic differences
become increasingly evident, reaching an amplitude of nearly $\Delta
z$/(1+$z$)=2$\times$10$^{-3}$ (600\kms) at redshifts $z$$\ga$4.

The origin of the redshift differences is primarily a 
systematic change in the ratio of the \ciii and \siiii emission lines in
the FD- and nFD-populations. The line ratio change results in a shift in
the centroid of the blended line; \siiii is weaker in the FD-detected quasars
and the blended line centroid moves redward. The \mgiicc redshifts for the FD 
quasars are thus too large.

\begin{figure}
 \begin{minipage}{\textwidth}
  \includegraphics[scale=0.35,angle=-90]{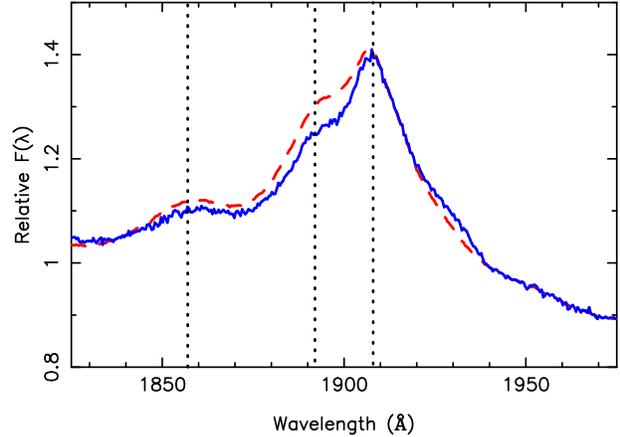}
  \end{minipage}
 \caption{Rest-frame spectra 
including the prominent emission line complex of \ciii
$\lambda$1908, \siiii $\lambda$1892 and \aliii $\lambda$1857, for the
master quasar template (red dash-dot line) and the FIRST-detected template
(blue solid line). The reference wavelengths of the three emission line
species are indicated by the vertical dotted lines. The significant 
difference in flux coincident with the location of the \siiii $\lambda$1892
is evident.}
\label{fig:rqrl_ciii}
\end{figure}

Based on the prescription of \citet{2007AJ....134..102S} for matching SDSS
quasars to the FIRST survey there are 4326 FD-quasars with $z$$>$1.1 in
the DR6 quasar catalogue.  The small fraction (7 per cent) of FD-quasars,
combined with the low amplitude of the systematic redshift differences,
means that the inclusion of FD-detected quasars in construction of the
master template results in changes to cross-correlation redshifts of
$\Delta z$/(1+$z$)$<$10$^{-4}$.  However, generation of an individual
template for the FD-quasars results in a quasar template with significant
differences in the form of the \ciii+\siiii+\aliii emission line
complex (Fig.~\ref{fig:rqrl_ciii}).

Construction of the new FD-quasar template proceeds in an identical fashion to
that described in Section~\ref{sec:temcon} but the individual quasars used
differ.  For redshifts $z$$<$1.0 the same quasars used to generate the master
quasar template are employed, whereas, at redshifts $z$$>1.0$, only FD-quasars
are used, with the minimum number of spectra required to generate a composite in
a redshift slice reduced from 100 to 50 (Section~\ref{sec:temcon}). The number
of quasars contributing at each redshift interval is given in Col.  4 of
Table~\ref{tab:cc_construct}. As evident from Fig.~\ref{fig:rqrl_ciii}, the
form of the \ciii+\siiii+\aliii emission line complex differs between
the master- and FD-quasar-templates. The size of the empirical transformations
necessary to bring the FD-quasar redshift estimates onto the reference system, in
which the \mgii emission line centroid does not vary, with either redshift or
absolute magnitude, are significantly reduced compared to the quasar 
population as a whole. 

For redshifts 1.1$\le$$z$$<$2.1 a reduction of $\Delta
z/(1+z)$=1.42$\times$10$^{-4}$ in the \mgiicc redshifts, independent of redshift
and absolute magnitude, is necessary.  For \ciiicc redshifts an absolute
magnitude dependent correction of 4.19$\times$10$^{-5}$, in the opposite
sense to that for the master template, is required, i.e.  for bright quasars
the \ciiicc redshifts are {\it reduced}.  However, given the $\simeq$4\,mag
dynamic range of the quasar sample, the maximum correction for any quasar is
$\Delta z/(1+z)$$\la$10$^{-4}$.  For the small number of quasars where it is
necessary to employ the \civ emission line, the \civcc redshifts require a
correction with slope 3.44$\times$10$^{-4}$, in the same sense as the larger
correction necessary for the master template, i.e.  for bright quasars
the \civcc redshifts are {\it increased} (Fig.~\ref{ciii_civ_raw}).

\begin{table*}
\begin{center}
\caption{\label{tab:cat} Quasar redshift catalogue.
The full table is available in the electronic version of the journal.}

\begin{tabular}{crrccrcrcrc} 
\hline
Name & RA & Dec & $z$ & $\sigma_z$ & FIRST & $z$ Alternate & Plate & MJD & Fibre & Code \\
     &  J2000 (deg)    & J2000 (deg) &          &       & &     &       &     &       &      \\
\hline
SDSS J000006.53+003055.2 &    0.02723 &    0.51534 &  1.823154 &  0.001025 &  -1 &  -999.0 &   685 &  52203 &  467 &  3 \\ 
SDSS J000008.13+001634.6 &    0.03390 &    0.27630 &  1.836332 &  0.000614 &  -1 &  -999.0 &   685 &  52203 &  470 &  3 \\ 
SDSS J000009.26+151754.5 &    0.03860 &   15.29848 &  1.197436 &  0.000369 &   0 &     1.200035 &   751 &  52251 &  354 &  2 \\ 
SDSS J000009.38+135618.4 &    0.03909 &   13.93845 &  2.240486 &  0.001468 &   0 &     2.240486 &   750 &  52235 &   82 &  7 \\ 
SDSS J000009.42-102751.9 &    0.03927 &  -10.46443 &  1.851731 &  0.000966 &  -1 &  -999.0 &   650 &  52143 &  199 &  3 \\ 
SDSS J000011.41+145545.6 &    0.04755 &   14.92935 &  0.460127 &  0.000357 &   0 &  -999.0 &   750 &  52235 &  499 &  1 \\ 
SDSS J000011.96+000225.3 &    0.04984 &    0.04036 &  0.478321 &  0.000358 &  -1 &  -999.0 &   387 &  51791 &  200 &  1 \\ 
SDSS J000012.25-003220.5 &    0.05108 &   -0.53905 &  1.437047 &  0.000697 &  -1 &  -999.0 &  1091 &  52902 &  129 &  3 \\ 
SDSS J000013.14+141034.6 &    0.05479 &   14.17630 &  0.949947 &  0.000521 &   0 &  -999.0 &   750 &  52235 &   98 &  3 \\ 
SDSS J000013.80-005446.8 &    0.05751 &   -0.91300 &  1.840606 &  0.000789 &  -1 &  -999.0 &  1091 &  52902 &  108 &  3 \\ 

\hline
\end{tabular}
\vspace*{-0.4cm}
\end{center}
\end{table*}

\begin{table*}
\begin{center}
\caption{\label{tab:spec} Quasar cross-correlation templates.
The full table is available in the electronic version of the journal.}
\begin{minipage}{14.5cm}
\begin{tabular}{ccccc} 
\hline
Wavelength & Relative Flux\footnote{Data for master template}         
& Number & Relative Flux\footnote{Data for FD-quasar template}         & Number \\
(\AA)      & (per unit wavelength) &        & (per unit wavelength) & \\
\hline
1275.26 & 8.130 & 179 & -999.0 & 0 \\
1275.56 & 7.582 & 186 & -999.0 & 0 \\
1275.85 & 7.845 & 191 & -999.0 & 0 \\
1276.15 & 7.780 & 196 & -999.0 & 0 \\
1276.44 & 7.673 & 204 & -999.0 & 0 \\
1276.73 & 7.682 & 212 & -999.0 & 0 \\
1277.03 & 7.730 & 219 & -999.0 & 0 \\
1277.32 & 7.862 & 229 & -999.0 & 0 \\
1277.61 & 7.644 & 234 & -999.0 & 0 \\
1277.91 & 7.755 & 247 & -999.0 & 0 \\
\hline
\end{tabular}
\end{minipage}
\end{center}
\vspace*{-0.4cm}
\end{table*}

\section{The Redshift Catalogue}

Table~\ref{tab:cat} includes redshifts and error estimates for 91\,665 quasars.
Col.  1:  is the SDSS coordinate object name, taken from the SDSS DR7 Legacy
Release whenever available.  Cols.  2 and 3 give the object J2000 right
ascension and declination in decimal degrees.  The redshift and redshift error
are given in Cols.  4 and 5 respectively.  Col.  6 provides a code specifying
the FIRST-detection (FD) status of the quasar (--1:  not detected, 0:  outside
FIRST footprint, 1:  detected).  Col.  7, alternate
redshift for quasars with Col.  6 detection code =0 (derived using the FD-quasar
template) and =1 (derived using the master quasar template).  The alternate
redshift is assigned a value of `--999.0' for quasars with Col.  6 detection
code =--1.  Col.  8 specifies the origin of the redshift estimate via a
numerical code (1:  \oiii, 2:  \oii, 3:  \mgiicc, 4:  \ciiicc, 5:  \civcc, 6:
extra\_cc, 7:  SDSS).  The SDSS spectrum from which the redshift estimate is
derived is specified via the spectroscopic plate number.  modified Julian date
of observation and fibre number in Cols.  9, 10 and 11 respectively.  The
redshifts are given to six decimal places but, as evident from the size of the
associated errors, the accuracy for individual objects is two orders of
magnitude larger.  The high level of precision is retained to avoid quantisation
when comparing different redshift estimates specified to only four decimal
places.

The provision of alternate redshifts for FD-quasars and quasars whose FIRST
detection status is unclear, allows the use of an appropriate redshift by
researchers with particular definitions of `radio'-quasar subsamples and/or
additional radio-observations for quasars outside the current FIRST footprint.
The primary (Col. 4) and alternate (Col. 7) redshifts differ only when the
primary redshift is derived from cross-correlation (with one of
the two quasar templates) and has a value $z$$>$1.1.

Two quasars, SDSS\,J134415.75+331719.1 and SDSS\,J142507.32+323137.4, exhibit
distinctive double-peaked narrow emission.  In both cases, the redshift
corresponding to the higher velocity system is included in the table.

The majority of researchers will be interested in the combined redshift
error (Col. 5) arising from the limited SNR of the SDSS spectra and intrinsic
variation from quasar-to-quasar. However, the internal
contribution can be recovered straightforwardly via use of the amplitude of the
quasar-to-quasar errors listed in Table~\ref{tab:z_error}. 

Table~\ref{tab:spec} presents the master quasar templates used to estimate the
cross-correlation redshifts.  Col.  1 lists the rest-frame wavelength (\AA).
Cols.  2, and 3 include the relative flux (per unit wavelength) and the number
of spectra contributing for the master template, while Cols.  4 and 5 provide
the same information for the FD-quasar template.  The FD-quasar template does
not extend quite as far to the blue and the flux column contains entries of
`--999.0' for wavelengths $\lambda$$<$1296.2\,\AA.

While the spectra should prove of use in the context of redshift estimation, the
templates are {\it not} suitable for studies of quasar spectral energy
distributions, where care must be taken in defining the large-scale shape of
such composite spectra.

\section{Discussion}

The approach adopted in this paper to the question of deriving redshifts with a
common zero-point over an extended dynamic range in redshift, and hence
involving disjoint spectral wavelength coverage, differs from that normally
employed.  The majority of studies to date have focussed on the parameterisation
of the rest-frame centroid differences between the strongest emission lines
present in the quasar spectra \citep[e.g.  Appendix A of][for a recent
example]{2007AJ....133.2222S}.  Use of the cross-correlation redshifts directly,
bypasses many of the difficulties associated in providing reliable,
reproducible, parameterisations of low SNR, asymmetric, often blended, emission
lines present on `continua' that also show significant variation from quasar to
quasar.  The resultant quasar-to-quasar dispersion and the internal
reproducibility of the new HW-redshifts represent significant improvements over
even the most careful studies utilising individual emission features.

Systematic, luminosity-dependent relative emission line shifts have not 
featured in many previous studies of the quasar population. In part, the
lack of such work may reflect the difficulty of performing such studies
prior to the availability of the more recent SDSS Data Releases. An exception
is the work of \citet{2002AJ....124....1R} who find a clear relationship
between emission line centroid shifts, of exactly the type discussed here,
and emission line equivalent width. \citet{2002AJ....124....1R} note that
the line equivalent width is directly related to quasar absolute magnitude
via the Baldwin Effect \citep{1977ApJ...214..679B}.

\begin{figure}
 \begin{minipage}{\textwidth}
\includegraphics[scale=0.30,angle=-90]{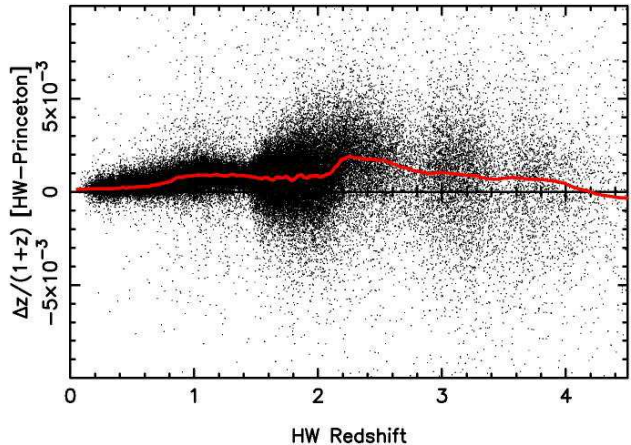}
  \end{minipage}
  
\caption{Redshift differences,$\Delta z/(1+z)$, between the final HW-redshifts
and Princeton redshifts, as a function of redshift.  Note the large range on the
y-axis. Data for 90\,979 quasars are
included (1306 lie outside the y-axis range plotted).}
\label{fig:hw_prince}
\end{figure}

\begin{figure*}
 \begin{minipage}{14cm}
\begin{center}
  \includegraphics[scale=0.6]{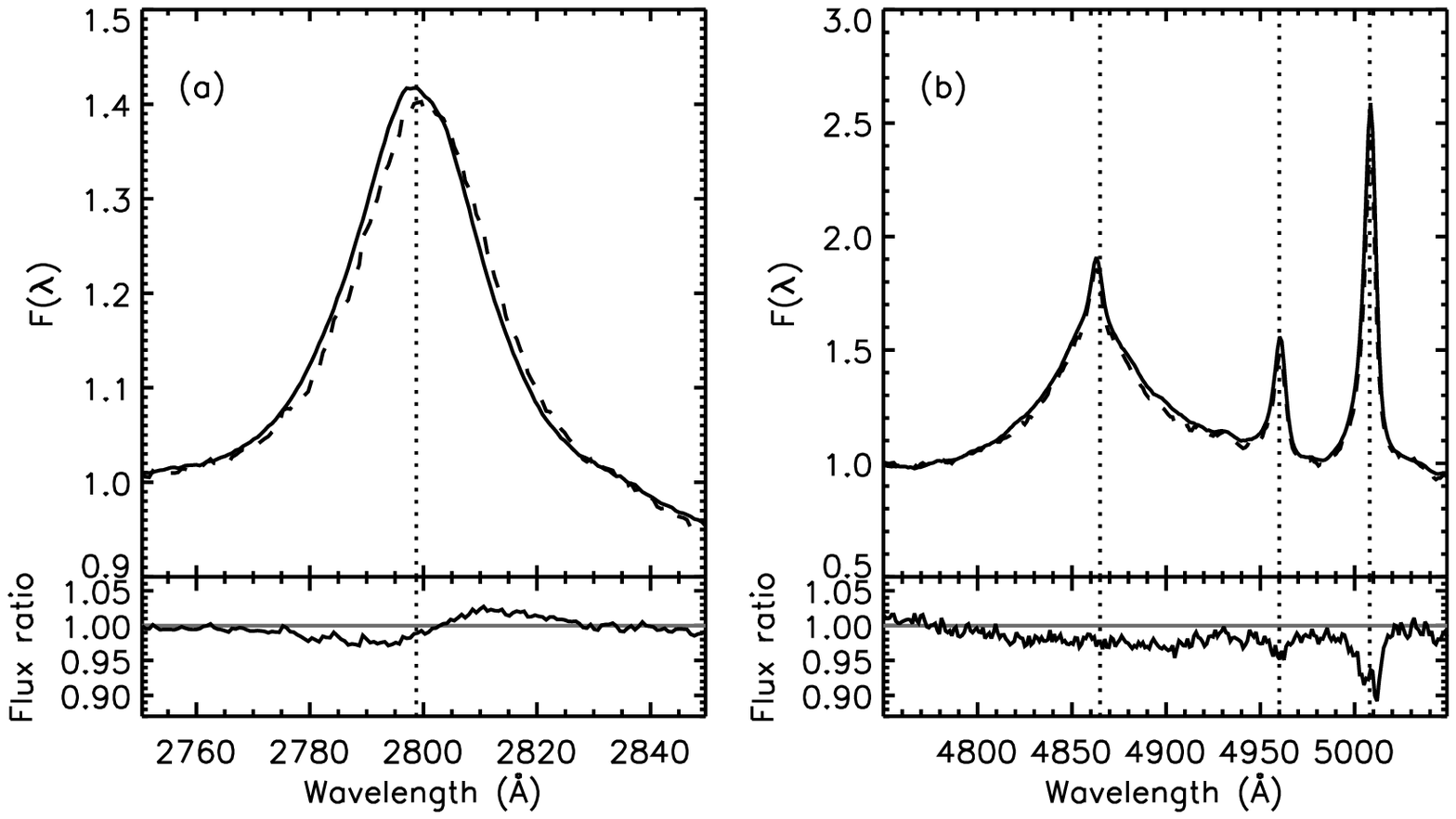}
\end{center}
  \end{minipage}
 \caption{Rest-frame spectra of the new quasar composite compared to that of
 \citet{2001AJ....122..549V}. Panel (a) shows
the prominent emission line of 
\mgii $\lambda\lambda$2796.35,2803.53, with the new quasar composite (solid line)
and the \citet{2001AJ....122..549V} composite (dashed line). The lower
plot shows the ratio \citet{2001AJ....122..549V}/new. The vertical dotted line
indicates the wavelength 2798.75\,\AA, derived from the \mgii components in
the ratio 2:1. The significantly bluer location of \mgii emission in the new
composite is evident. Panel (b) shows the same information for the rest-frame
wavelength region including H$\beta$ and \oiii $\lambda\lambda$4960,5008. 
The vertical dotted lines indicate the rest-frame wavelengths of 4862.68
4960.30 and 5008.24\,\AA \ for H$\beta$ and \oiii. 
While the new composite possesses slightly stronger \oiii emission there is no
evidence for any detectable offset in the emission line locations of 
H$\beta$ or \oiii.}
\label{fig:comp_comp}
\end{figure*}

\subsection{ Comparison with Princeton redshifts and the 
\citet{2001AJ....122..549V} quasar template}

Operationally, it is found that straightforward systematic corrections to quasar
redshift estimates, as a function of quasar absolute magnitude, reduce the
systematic trends as a function of redshift (absolute magnitude) to $<$30\kms
per unit redshift ($<$10\kms per magnitude).  Internal reproducibility
represents a factor $>$2 improvement over the SDSS redshift determinations.  The
results presented above, combined with the form of the differences between the
Princeton and SDSS redshifts (Section~\ref{sec:sdss_zs}), show that the origin
of a significant proportion of the improvements achieved are due to differences
in the cross-correlation procedure/algorithm employed.  However, comparison of
the new HW-redshifts with the Princeton determinations
(Fig.~\ref{fig:hw_prince}) still shows large ($\simeq$600\kms at $z$$\simeq$2.2)
systematic differences.

The two evident `jumps' in the relationship between the HW- and SDSS- estimates
occur as cross-correlation redshifts including the \mgii $\lambda$2799 emission
line become important (at $z$$\sim$0.8) and where \mgii moves beyond the red
limit of the SDSS spectra (at $z$$\simeq$2.1).  The behaviour can be traced
directly to differences in the new quasar template and that of
\citet{2001AJ....122..549V}.  Fig.~\ref{fig:comp_comp}b shows the excellent
agreement between the composites at optical wavelengths where emission lines,
including H$\beta$ and \oiii $\lambda\lambda$4960,5008, dominate the redshift
determinations (either directly, via emission line locations, or, through the
contribution of emission lines to the cross-correlation signal).  In contrast,
Fig.~\ref{fig:comp_comp}a, illustrates the significant difference in the
location of the \mgii $\lambda$2799 emission line in the two composites.  At the
accuracy levels of interest, absolute wavelength `centroids' of broad emission
lines in quasar spectra are dominated by the particular scheme used to define
the associated `continuum' and the height above the continuum used to define the
`line'.  However, the centroid of the portion of the \mgii emission line above
half the peak height is 1.2$\pm$0.1\,\AA \ bluer in the HW-template compared to
the \citet{2001AJ....122..549V} template.  The line centroid moves redward as
increasingly large fractions of the line wings are included but at the half peak
height level the HW-template line centroid is only 0.4$\pm0.1$\,\AA \
($\simeq$45\kms) redward of the rest-frame reference value of 2798.75\,\AA,
derived from the \mgii components in the ratio 2:1.

The strongest `jump' in the relation between the HW- and Princeton-redshifts in
Fig.~\ref{fig:hw_prince}, at $z$$\simeq$2.2, derives fundamentally from the
large systematic trends in the ratio of \mgii to `\ciii' emission line locations
as a function of quasar absolute magnitude (Fig.~\ref{fig:mg_ciii}).  The origin
of the effect is primarily the change in the ratio of \ciii $\lambda$1908 to
\siiii $\lambda$1892 (see Fig.~\ref{fig:rqrl_ciii} for illustration in the
context of FIRST-detected quasars).  A direct comparison of the HW-template and
\citet{2001AJ....122..549V} template is somewhat misleading because the
HW-redshifts are derived only following the significant absolute magnitude
dependent corrections.  However, using any sensible definition of the emission
line, the \ciii + \siiii blend is significantly bluer in the HW-template than
in the \citet{2001AJ....122..549V} composite, producing the increase
in the HW-redshifts at $z$$\simeq$2.2.  The Princeton redshifts then become
progressively closer to the HW-redshifts as the \civ emission line (with its
well-established increasing blue asymmetry at increasing quasar luminosity)
dominates the Princeton determinations at higher redshifts.  Recall though, that
the \civ emission region does {\it not} contribute to the HW-redshifts, except
in a very small percentage of quasars.

\subsection{Associated \civ and \mgii absorbers as quasar redshift diagnostics}

The availability of the large SDSS quasar catalogues have stimulated new
investigations into the physical origin of associated absorbers, particularly
those evident through the presence of \civ and \mgii absorption
\citep[e.g.][]{2008MNRAS.386.2055N, 2008MNRAS.388..227W, 2008ApJ...679..239V}.
A pre-requisite for such investigations is an estimate of the systemic quasar
redshifts.  Given the large intrinsic variation in the properties of the \civ
emission line and the relative invariance of the \mgii emission line centroid,
redshifts based on the location of the \mgii emission line are often employed in
studies of both associated \civ and \mgii absorbers in quasars with redshifts
$z$$\le$2.1.

\begin{figure}
 \begin{minipage}{\textwidth}
  \includegraphics[scale=0.35,angle=-90]{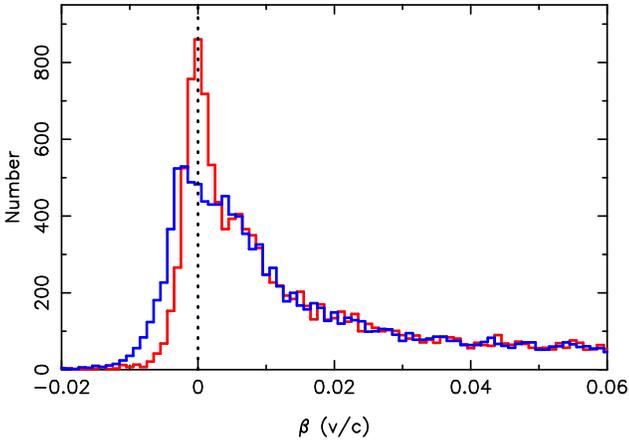}
  \end{minipage}
  \caption{Observed frequency distribution of redshift differences,
$\beta$=$v/c$, for $\simeq$23\,800 \civ absorbers using both SDSS- (blue) and
HW-redshifts (red) for quasars with redshifts 1.55$<$$z$$<$3.5.  The
HW-redshift-based histogram shows a much higher, better defined, peak, centred
close to $\beta$$\simeq$0 and a greatly reduced population of absorbers with
positive $\beta$ values (i.e.  $z_{abs}$$>$$z_{qso}$).  The centroid of the
$\beta$$\simeq$0 component for the HW-redshifts shows no detectable shift over
the entire redshift range of the quasars, 1.55$<$$z$$<$3.5.  }
\label{fig:abs_civ}
\end{figure}

\begin{figure}
 \begin{minipage}{\textwidth}
  \includegraphics[scale=0.35,angle=-90]{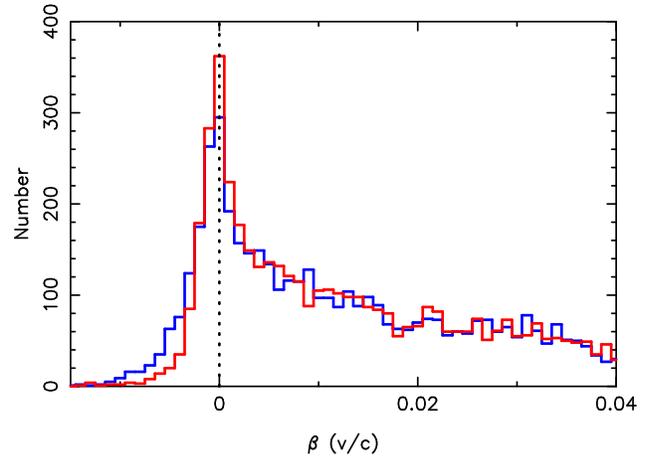}
  \end{minipage}
  \caption{Observed frequency distribution of redshift differences,
$\beta$=$v/c$, for $\simeq$8750 \mgii absorbers using both SDSS- (blue) and
HW-redshifts (red) for quasars with redshifts 0.45$<$$z$$<$2.1.  The
HW-redshift-based histogram shows a higher, better defined, peak, centred close
to $\beta$$\simeq$0 and a significantly reduced population of absorbers with
positive $\beta$ values (i.e.  $z_{abs}$$>$$z_{qso}$).  The centroid of the
$\beta$$\simeq$0 component for the HW-redshifts shows no detectable shift over
the entire redshift range of the quasars, 0.45$<$$z$$<$2.1.}
\label{fig:abs_mgii}
\end{figure}

Both \mgii and \civ absorber catalogues are available from our investigation of
absorber populations in SDSS quasars \citep[e.g.][]{2006MNRAS.367..211W}.
Strong narrow absorbers are flagged and `removed' from the quasar spectra prior
to the calculation of the cross-correlation redshift determinations
(Section~\ref{sec:temcon}).  The new HW-quasar redshifts are thus essentially
independent of the presence of individual absorbers and a comparison of
associated absorbers velocity distributions, using both SDSS- and HW-redshifts,
provides a powerful test of the redshift accuracy in an astrophysical context of
considerable current interest.  Fig.~\ref{fig:abs_civ} shows the observed
distribution of redshift differences \footnote{The observed distributions are
shown.  No attempt has been made to calculate an absorber density by
incorporating the redshift-path accessible as a function of $\beta$.},
$\beta$=$v/c$, for $\simeq$23\,800 \civ absorbers using both SDSS- and
HW-redshifts for quasars with redshifts 1.55$<$$z$$<$3.5.  The differences in
the distributions are striking, with the HW-redshift-based histogram showing a
much higher peak at $\beta$$\simeq$0 and a greatly reduced population of
positive $\beta$ values (i.e.  $z_{abs}$$>$$z_{qso}$).  The centroid of the
$\beta$$\simeq$0 component shows no detectable shift over the entire redshift
range of the quasars, 1.55$<$$z$$<$3.5.

Fig.~\ref{fig:abs_mgii} shows the equivalent distribution of redshift 
differences, for $\simeq$8750 \mgii absorbers using both SDSS- and 
HW-redshifts for quasars with redshifts 0.45$<$$z$$<$2.1. The differences
between the SDSS- and HW-redshifts at $z$$<$2.1 are significantly smaller
than for the quasars included in the \civ absorber sample. However, 
similar behaviour is evident to that seen in the \civ absorbers, the 
$\beta$$\simeq$0 peak is significantly better defined using the 
HW-redshifts and the population of redshifted absorbers with positive 
$\beta$ greatly reduced. A more detailed consideration of the distribution
of the \mgii absorber redshifts leads to an improved parameterisation of
the various constituent absorber populations \citep{2009arXiv0907.5221W}.
 
To summarise, the distributions of the \civ and \mgii associated absorber
populations provide independent confirmation of the veracity of the new
HW-redshifts. While beyond the scope of the present paper, investigations 
of associated absorber populations are in hand using the full SDSS DR7
Legacy Release spectroscopic database.
 
\section{conclusions}

A systematic investigation of the relationship between different redshift
estimation schemes for more than 91\,000 quasars in the Sloan Digital Sky Survey
(SDSS) Data Release 6 (DR6) is presented.  Empirical relationships between
redshifts based on i) \caii H \& K host galaxy absorption, ii) quasar \oii
$\lambda$3728, iii) \oiii $\lambda\lambda$4960,5008 emission, and iv)
cross-correlation (with a master quasar template) that includes, at increasing
quasar redshift, the prominent \mgii $\lambda$2799, \ciii $\lambda$1908 and \civ
$\lambda$1549 emission lines, are established as a function of quasar redshift
and luminosity.  New redshifts in the resulting catalogue possess systematic
biases a factor of $\simeq$20 lower compared to the SDSS redshift values;
systematic effects are reduced to the level of $\Delta z/(1+z)$$\le$10$^{-4}$
(30\kms) per unit redshift, or $\le$2.5$\times$10$^{-5}$ per unit absolute
magnitude.

It is important to realise that there will be systematic redshift trends present
as a function of the quasar SEDs and the specific example of FIRST-detected
quasars (Section~\ref{radio_qsos}) provides an example, related to the
radio-properties of the quasar SEDs.  One of the primary motivations of this
work is to facilitate further studies of SED-dependent systematic emission line
properties, working from redshift estimates whose properties as a function of
redshift and absolute magnitude are well understood.

Equally important as the new redshift determinations, well-determined empirical
estimates of the quasar-to-quasar dispersion in redshifts are available for each
method of redshift estimation and a combined internal+population uncertainty is
provided for every quasar in the catalogue.

The improved redshifts and their associated errors have wide applicability in
areas such as quasar absorption outflows, quasar clustering, quasar-galaxy
clustering and proximity-effect determinations.

\section*{acknowledgments}

We thank the referee, Don Schneider, for provided a very careful and
constructive reading of the original draft.  We are grateful to James Allen, Bob
Carswell and Gordon Richards for encouragement, insights and helpful
conversations.  PCH acknowledges support from the STFC-funded Galaxy Formation
and Evolution programme at the Institute of Astronomy.  VW is supported by a
Marie Curie Intra-European Fellowship.

Funding for the SDSS and SDSS-II has been provided by the Alfred P.  Sloan
Foundation, the Participating Institutions, the National Science Foundation, the
U.S.  Department of Energy, the National Aeronautics and Space Administration,
the Japanese Monbukagakusho, the Max Planck Society, and the Higher Education
Funding Council for England.  The SDSS Web Site is http://www.sdss.org/.

The SDSS is managed by the Astrophysical Research Consortium for the
Participating Institutions.  The participating institutions are the American
Museum of Natural History, Astrophysical Institute Potsdam, University of Basel,
University of Cambridge, Case Western Reserve University, University of Chicago,
Drexel University, Fermilab, the Institute for Advanced Study, the Japan
Participation Group, Johns Hopkins University, the Joint Institute for Nuclear
Astrophysics, the Kavli Institute for Particle Astrophysics and Cosmology, the
Korean Scientist Group, the Chinese Academy of Sciences (LAMOST), Los Alamos
National Laboratory, the Max-Planck-Institute for Astronomy (MPIA), the
Max-Planck-Institute for Astrophysics (MPA), New Mexico State University, Ohio
State University, University of Pittsburgh, University of Portsmouth, Princeton
University, the United States Naval Observatory, and the University of
Washington.

\bibliographystyle{mn2e}
% Use this when working
%\bibliography{refs_all}

\begin{thebibliography}{}

\bibitem[\protect\citeauthoryear{Abazajian et 
al.}{2009}]{2009ApJS..182..543A} Abazajian K.~N., et al., 2009, ApJS, 182, 
543 

\bibitem[\protect\citeauthoryear{Adelman-McCarthy et 
al.}{2007}]{2007ApJS..172..634A} Adelman-McCarthy J.~K., et al., 2007, 
ApJS, 172, 634 

\bibitem[\protect\citeauthoryear{Adelman-McCarthy et 
al.}{2008}]{2008ApJS..175..297A} Adelman-McCarthy J.~K., et al., 2008, 
ApJS, 175, 297 

\bibitem[\protect\citeauthoryear{Bajtlik, Duncan, 
\& Ostriker}{1988}]{1988ApJ...327..570B} Bajtlik S., Duncan R.~C., Ostriker J.~P., 1988, ApJ, 327, 570 

\bibitem[\protect\citeauthoryear{Baldwin}{1977}]{1977ApJ...214..679B} 
Baldwin J.~A., 1977, ApJ, 214, 679 

\bibitem[\protect\citeauthoryear{Becker, White, 
\& Helfand}{1995}]{1995ApJ...450..559B} Becker R.~H., White R.~L., Helfand D.~J., 1995, ApJ, 450, 559 

\bibitem[\protect\citeauthoryear{Boroson}{2005}]{2005AJ....130..381B}
Boroson T., 2005, AJ, 130, 381

\bibitem[\protect\citeauthoryear{Croom et al.}{2002}]{2002MNRAS.335..459C} 
Croom S.~M., Boyle B.~J., Loaring N.~S., Miller L., Outram P.~J., Shanks 
T., Smith R.~J., 2002, MNRAS, 335, 459 

\bibitem[\protect\citeauthoryear{Gaskell}{1982}]{1982ApJ...263...79G} 
Gaskell C.~M., 1982, ApJ, 263, 79 

\bibitem[\protect\citeauthoryear{Gibson et al.}{2009}]{2009ApJ...692..758G} 
Gibson R.~R., et al., 2009, ApJ, 692, 758 

\bibitem[\protect\citeauthoryear{Heckman et 
al.}{1981}]{1981ApJ...247..403H} Heckman T.~M., Miley G.~K., van Breugel 
W.~J.~M., Butcher H.~R., 1981, ApJ, 247, 403 

\bibitem[\protect\citeauthoryear{{Hewett}, {Irwin}, {Bunclark}, {Bridgeland},
  {Kibblewhite}, {He} \& {Smith}}{{Hewett} et~al.}{1985}]{1985MNRAS.213..971H}
{Hewett} P.~C.,  {Irwin} M.~J.,  {Bunclark} P.,  {Bridgeland} M.~T.,
  {Kibblewhite} E.~J.,  {He} X.~T.,    {Smith} M.~G.,  1985, \mnras, 213, 971
  
\bibitem[\protect\citeauthoryear{Kirkman 
\& Tytler}{2008}]{2008MNRAS.391.1457K} Kirkman D., Tytler D., 2008, MNRAS, 391, 1457 

\bibitem[\protect\citeauthoryear{Nestor, Hamann, 
\& Hidalgo}{2008}]{2008MNRAS.386.2055N} Nestor D., Hamann F., Hidalgo P.~R., 2008, MNRAS, 386, 2055 

\bibitem[\protect\citeauthoryear{Padmanabhan et 
al.}{2009}]{2009MNRAS.397.1862P} Padmanabhan N., White M., Norberg P., 
Porciani C., 2009, MNRAS, 397, 1862 

\bibitem[\protect\citeauthoryear{Richards}{2006}]{2006astro.ph..3827R} 
Richards G.~T., 2006, arXiv:astro-ph/0603827 

\bibitem[\protect\citeauthoryear{Richards et 
al.}{2002}]{2002AJ....124....1R} Richards G.~T., Vanden Berk D.~E., 
Reichard T.~A., Hall P.~B., Schneider D.~P., SubbaRao M., Thakar A.~R., 
York D.~G., 2002, AJ, 124, 1 

\bibitem[\protect\citeauthoryear{Schneider et 
al.}{2007}]{2007AJ....134..102S} Schneider D.~P., et al., 2007, AJ, 134, 
102 

\bibitem[\protect\citeauthoryear{Schneider et 
al.}{2010}]{2010AJ....XXX..XXXX} Schneider D.~P., et al., 2010, AJ, in press

\bibitem[\protect\citeauthoryear{Shen et al.}{2007}]{2007AJ....133.2222S} 
Shen Y., et al., 2007, AJ, 133, 2222 

\bibitem[\protect\citeauthoryear{Stoughton et 
al.}{2002}]{2002AJ....123..485S} Stoughton C., et al., 2002, AJ, 123, 485 

\bibitem[\protect\citeauthoryear{Tonry 
\& Davis}{1979}]{1979AJ.....84.1511T} Tonry J., Davis M., 1979, AJ, 84, 1511 

\bibitem[\protect\citeauthoryear{Tytler 
\& Fan}{1992}]{1992ApJS...79....1T} Tytler D., Fan X.-M., 1992, ApJS, 79, 1 

\bibitem[\protect\citeauthoryear{Tytler et al.}{2009}]{2009MNRAS.392.1539T} 
Tytler D., et al., 2009, MNRAS, 392, 1539 

\bibitem[\protect\citeauthoryear{Vanden Berk et 
al.}{2001}]{2001AJ....122..549V} Vanden Berk D.~E., et al., 2001, AJ, 122, 
549 

\bibitem[\protect\citeauthoryear{Vanden Berk et 
al.}{2008}]{2008ApJ...679..239V} Vanden Berk D., et al., 2008, ApJ, 679, 
239 

\bibitem[\protect\citeauthoryear{Wild}{2009}]{2009arXiv0907.5221W} Wild V., 
2009, arXiv:0907.5221 

\bibitem[\protect\citeauthoryear{Wild 
\& Hewett}{2005}]{2005MNRAS.358.1083W} Wild V., Hewett P.~C., 2005, MNRAS, 358, 1083 

\bibitem[\protect\citeauthoryear{Wild, Hewett, 
\& Pettini}{2006}]{2006MNRAS.367..211W} Wild V., Hewett P.~C., Pettini M., 2006, MNRAS, 367, 211 

\bibitem[\protect\citeauthoryear{Wild et al.}{2008}]{2008MNRAS.388..227W} 
Wild V., et al., 2008, MNRAS, 388, 227 

\bibitem[\protect\citeauthoryear{York et al.}{2000}]{2000AJ....120.1579Y} 
York D.~G., et al., 2000, AJ, 120, 1579 

\end{thebibliography}

\end{document}